\documentclass{ieeeaccess}
\usepackage{graphicx}
\usepackage{amsmath, amssymb, amsfonts, amsthm, mathtools, bm, gensymb}
\usepackage{url}
\usepackage{caption}
\usepackage{siunitx}
\usepackage{algorithm, algpseudocode}
\usepackage{cite}
\usepackage{textcomp}
\usepackage{tikz, pgfplots}

\NewSpotColorSpace{PANTONE}
\AddSpotColor{PANTONE} {PANTONE3015C} {PANTONE\SpotSpace 3015\SpotSpace C} {1 0.3 0 0.2}
\SetPageColorSpace{PANTONE}%

\mathtoolsset{showonlyrefs} 
\usetikzlibrary{plotmarks,fit,shapes}
\usetikzlibrary{shapes,arrows}

\newcommand{\mc}[1]{\ensuremath{\mathcal{#1}}}
\newcommand{\mbb}[1]{\ensuremath{\mathbb{#1}}}
\newcommand{\tran}{\mathsf{T}}
\newcommand{\hermit}{\mathsf{H}}
\newcommand{\neuclid}[1]{\ensuremath{\left\|#1\right\|_2}}
\newcommand{\frob}[1]{\ensuremath{\left\|#1\right\|_\textrm{F}}}
\newcommand{\esp}[1]{\ensuremath{\mathbb{E}\left[#1\right]}}
\newcommand{\Rssu}{\ensuremath{\bm{R}_{s,u}}}

\newcommand{\Fu}{\ensuremath{\bm{F}_u}}
\newcommand{\Wu}{\ensuremath{\bm{W}_u}}
\newcommand{\Fiu}{\ensuremath{\bm{F}_{\text{i},u}}}
\newcommand{\Fij}{\ensuremath{\bm{F}_{\text{i},j}}}
\newcommand{\Fou}{\ensuremath{\bm{F}_{\text{o},u}}}
\newcommand{\Foj}{\bm{F}_{\text{o},j}}
\newcommand{\Wiu}{\ensuremath{\bm{W}_{\text{i},u}}}

\newcommand{\Wou}{\ensuremath{\bm{W}_{\text{o},u}}}
\newcommand{\Heffuj}{\ensuremath{\bm{H}_{\text{eff},u,j}}}

\newcommand{\Heffu}{\ensuremath{\bm{H}_{\text{eff},u}}}
\newcommand{\Cdl}{\ensuremath{\bm{C}_{\text{dl},u}}}
\newcommand{\Cul}{\ensuremath{\bm{C}_{\text{ul},u}}}
\DeclareMathOperator{\Blkdiag}{Blkdiag}
\DeclareMathOperator{\Diag}{Diag}
\DeclareMathOperator{\trace}{Tr}
\DeclareMathOperator{\rank}{rank}
\DeclareMathOperator{\vecspan}{span}

\DeclareMathOperator*{\argmax}{arg\,max}

\def\BibTeX{{\rm B\kern-.05em{\sc i\kern-.025em b}\kern-.08em
		T\kern-.1667em\lower.7ex\hbox{E}\kern-.125emX}}

\begin{document}

\history{Date of publication xxxx 00, 0000, date of current version xxxx 00, 0000.}
\doi{10.1109/ACCESS.2019.2949945}

\title{Double-Sided Massive MIMO Transceivers for MmWave Communications}
\author{\uppercase{Lucas N. Ribeiro}\authorrefmark{1},
		\uppercase{Stefan Schwarz}\authorrefmark{2}, \IEEEmembership{Senior Member,~IEEE}, and
		\uppercase{Andr\'e L. F. de Almeida}\authorrefmark{1}, \IEEEmembership{Senior Member,~IEEE}}

\address[1]{Wireless Telecommunications Research Group (GTEL), Universidade Federal do Cear\'a, Fortaleza CE 60440-900, Brazil}
\address[2]{Christian Doppler Laboratory for Dependable Wireless Connectivity for the Society in Motion, Institute of Telecommunications, Technische Universit\"{a}t Wien, 1040 Wien, Austria}

\tfootnote{The financial support by the Austrian Federal Ministry for Digital, and Economic Affairs and the National Foundation for Research, Technology and Development is gratefully acknowledged. This work was partially supported by the Brazilian National Council for Scientific and Technological Development - CNPq, CAPES/PROBRAL Proc. numbers 88887.144009/2017-00, 308317/2018-1, and FUNCAP.}

\markboth
{L. N. Ribeiro \headeretal: Double-Sided Massive MIMO Transceivers for MmWave Communications}
{L. N. Ribeiro \headeretal: Double-Sided Massive MIMO Transceivers for MmWave Communications}

\corresp{Corresponding author: Lucas N. Ribeiro (e-mail: nogueira@gtel.ufc.br).}

\begin{abstract}
	We propose practical transceiver structures for double-sided massive multiple-input-multiple-output (MIMO) systems. Unlike standard massive MIMO, both transmit and receive sides are equipped with high-dimensional antenna arrays. We leverage the multi-layer filtering architecture and propose novel layered transceiver schemes with practical channel state information requirements to simplify the complexity of our double-sided massive MIMO system. We conduct a comprehensive simulation campaign to investigate the performance of the proposed transceivers under different channel propagation conditions and to identify the most suitable strategy. Our results show that the covariance matrix eigenfilter design at the outer transceiver layer combined with maximum eigenmode transmission precoding/minimum mean square error combining at the inner transceiver layer yields the best achievable sum rate performance for different propagation conditions and multi-user interference levels.
\end{abstract}


\titlepgskip=-15pt

\maketitle

\section{Introduction} \label{sec:intro}

\PARstart{M}{assive} multiple-input multiple-output (MIMO) is one of the key technologies of modern mobile communication systems~\cite{bjornson2019massive,bjornson2019reality,sanguinetti2019towards}. It consists of employing a large number of antennas at the base station (BS) to provide a significant beamforming gain and to simultaneously serve several users. The canonical massive MIMO model~\cite{marzetta2016fundamentals} considers time division duplex (TDD) operation at sub-$6$ GHz frequencies, which allows for relatively simple channel state information (CSI) acquisition. The ever-increasing demand for system capacity and applicability in more general scenarios calls for novel massive MIMO extensions. For example, there are research efforts for developing novel massive MIMO techniques in different scenarios, including: frequency division duplex (FDD)~\cite{dai2018fdd}, cell-free systems~\cite{alonzo2019energy}, large intelligent surface aided MIMO~\cite{hu2018beyond}, and millimeter-wave (mmWave) systems~\cite{hemadeh2017millimeter,ribeiro2018energy,uwaechia2019spectral}.

MmWave massive MIMO has attracted much interest due to the promise of large available bandwidth and less strict regulation~\cite{hemadeh2017millimeter}. These features are crucial for novel application scenarios such as wireless backhauling~\cite{gao2015mmwave,siddique2015wireless,schwarz2018cellular} and vehicle-to-vehicle communications~\cite{choi2016millimeter}. However, mmWave systems face many propagation challenges such as atmospheric attenuation, strong free space loss, and material absorption~\cite{hemadeh2017millimeter}. Massive MIMO has been proposed to compensate for these issues with large beamforming gain. Most works, however, only consider users with a small number of antennas relative to the BS. Double-sided massive MIMO refers to the scenario wherein both BS and user equipment (UE) employ large antenna arrays. Therefore, this extension is even more suited than the standard massive MIMO implementation to operate at mmWave ranges, since it offers larger beamforming gain to offset the important signal propagation losses. Implementing this double-sided scenario in classical BS-smartphone links may not be realistic due to physical constraints in the latter. However, we can mention many application scenarios that may strongly benefit from this technology, including MIMO heterogeneous networks with wireless backhauling~\cite{ni2019enhancing}, terahertz communication systems~\cite{akyildiz2016realizing,nie2019intelligent,sarieddeen2019terahertz} and mmWave unmanned aerial vehicle communications~\cite{zhang2019research}.

Low-complexity transceivers for double-sided massive MIMO systems were first investigated in~\cite{schwarz2016performance}. The authors were interested in evaluating the effect of spatial antenna correlation on system performance. To this end, the Kronecker correlation model was adopted and the system performance was evaluated assuming linear transceiver schemes and perfect CSI. It was found that the impact of antenna correlation on performance strongly depends on the transceiver architecture. Specifically, zero-forcing (ZF) precoding and maximum eigenmode reception~(MER) showed robustness against strong antenna correlation provided that the number of served users is not as large as the number of BS antennas. Hybrid analog/digital (A/D) and fully-digital double-sided massive MIMO transceivers were investigated in~\cite{buzzi2018efficiency}. Partial ZF (PZF) and channel matching were proposed for both hybrid A/D and fully-digital strategies. However, it is not discussed whether the proposed transceiver architectures have practical CSI requirements. The transceiver strategies of \cite{schwarz2016performance} and~\cite{buzzi2018efficiency} rely on the perfect knowledge of the channel matrix of all users. As the size of these matrices is very large (due to the double-sided massive MIMO assumption), feedback and channel estimation techniques may become overwhelming.

A potential solution to the complexity of double-sided massive MIMO systems is multi-layer filtering~\cite{caire2013jsdm,ribeiro2017low,alkhateeb2017multilayer}. In this method, the filter matrix is decomposed as a product of lower-dimensional filter matrices, wherein each matrix (layer) is designed to achieve a single filtering task. The main motivation behind this idea is to enable efficient and low-complexity filtering in massive MIMO, which is challenging due to the large number of antennas. An attractive advantage of the multi-layer strategy is that, by decoupling the filter design problem for each layer, one can formulate simple sub-problems, which may be less computationally expensive than optimizing a large-dimensional full filter matrix. Another appealing feature is the successive dimensionality reduction. Each layer is associated with an effective channel matrix whose dimensions are smaller than those of the original channel. Therefore, the channel training overhead for these layers is reduced~\cite{alkhateeb2017multilayer}. The layered filter architecture allows designing each layer according to different CSI requirements. For example, in a two-layer approach, the first layer may depend on second-order channel statistics, while the second layer is based on the instantaneous knowledge of the low-dimensional effective channel generated by the composition of the first layer filters and the actual physical channel.

In~\cite{caire2013jsdm}, a two-layer joint spatial division and multiplexing (JSDM) filter is presented. The first layer consists of a pre-beamforming stage to group UEs with similar covariance eigenspace, while the second layer manages multi-user interference. We present a two-layer equalizer scheme for a single-user multi-stream massive MIMO system in~\cite{ribeiro2017low}. The first layer consists of a spatial ZF equalizer and the second layer is a low-dimensional minimum mean square error (MMSE) filter applied to the effective channel. We show that the proposed layered filtering approach is less complex than the standard MMSE equalizer since we decouple the filtering operation into simpler operations. In~\cite{schwarz2019grassmannian}, a novel Grassmannian product codebook scheme was proposed for limited feedback FDD massive MIMO systems with two-layer precoding filters. Analytical asymptotic approximations of the achievable transmission rate were obtained for the imperfect CSI scenario. In~\cite{alkhateeb2017multilayer}, the two-layer idea is generalized to the three-layer scenario: the first layer cancels the inter-cell interference, the second layer increases the desired signal power and the third layer mitigates intra-cell interference. The multi-layer framework of~\cite{alkhateeb2017multilayer} generalizes JSDM to also suppress inter-cell interference. The multi-layer strategy was recently applied to a cloud radio access network using full-dimension MIMO in~\cite{femenias2018multi} and novel precoding schemes combined with the multi-layer strategy were also presented in~\cite{rezaei2019multi}.

The main contributions of the present work are:
\begin{itemize}
	\item We propose low-complexity multi-layer double-sided massive MIMO transceivers with practical CSI requirements;
	\item We provide a novel outer layer filter design method based on partial CSI knowledge, herein referred to as semi-orthogonal path selection;
	\item We conduct a comprehensive simulation-based study of several double-sided massive MIMO transceivers, including the proposed ones. We also conduct benchmark simulations to discuss the advantages of the proposed methods;
	\item We discuss the applicability of the presented methods for different mmWave channel setups and indicate the propagation conditions where multiple data stream transmission per UE is feasible.
\end{itemize}

We provide the signal, system and channel models as well as details on CSI acquisition in Section~\ref{sec:sysmodel}. We introduce our transceiver schemes and discuss their computational complexity in Section~\ref{sec:trx}. We present our simulation results and discussions in Section~\ref{sec:sim} and we conclude our paper in Section~\ref{sec:conclusion}.

\begin{figure*}[ht]
	\centering
	\includegraphics[width=\textwidth]{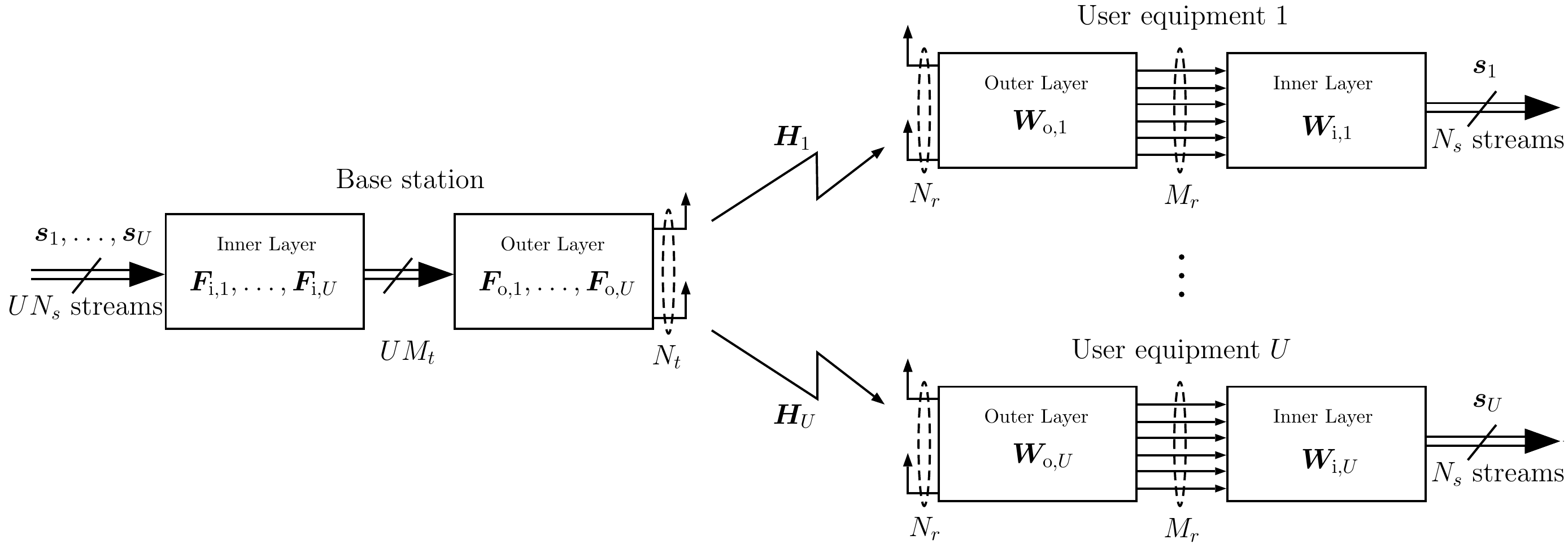}
	\caption{Illustration of the considered multi-layer double-sided massive MIMO system model.}
	\label{fig:model}
\end{figure*}

\emph{Notation:} Vectors and matrices are written as lowercase and uppercase boldface letters, respectively, e.g., $\bm{x}$ and $\bm{X}$. The $(i,j)$-th entry of $\bm{X}$ is written as $[\bm{X}]_{i,j}$. The transpose and the conjugate transpose (Hermitian) of $\bm{X}$ are represented by $\bm{X}^\tran$ and $\bm{X}^\hermit$, respectively. The $N$-dimensional identity matrix is represented by $\bm{I}_{N}$ and the $(M\times N)$-dimensional null matrix by $\bm{0}_{M\times N}$. The imaginary unit is referred to as $\jmath = \sqrt{-1}$. The Euclidean norm, the Frobenius norm, the matrix trace, the determinant, and the statistical expected value are respectively denoted by $\neuclid{\cdot}$, $\frob{\cdot}$, $\trace(\cdot)$, $\det(\cdot)$, and $\esp{\cdot}$. The $\Diag(\cdot)$ operator transforms an input vector into a diagonal matrix and $\Blkdiag(\cdot)$ forms a block-diagonal matrix from the matrix inputs. The operator $\rank(\cdot)$ denotes the argument matrix's rank, $\vecspan(\cdot)$ refers to the  space spanned by the argument vectors, and $\#(\cdot)$ denotes the argument set's cardinality. The uniform distribution from $a$ to $b$ is denoted $\mc{U}(a,b)$. The complex Gaussian distribution with mean $\bm{\mu}$ and covariance matrix $\bm{\Sigma}$ is written as $\mc{CN}(\bm{\mu}, \bm{\Sigma})$. $O(\cdot)$ stands for the Big-O complexity notation and $\stackrel{!}{=}$ denotes equality by construction.

\section{System Model} \label{sec:sysmodel}

Let us consider the single-cell multi-user MIMO system depicted in Figure~\ref{fig:model}. Assuming downlink operation, a single base station equipped with $N_t$ antennas communicates with $U$ UEs, each having $N_r$ antennas. We assume the \emph{double-sided massive} scenario, i.e., the BS and UEs are equipped with a large number ($\geq 64$) of antennas. We consider multi-stream transmission: the BS sends $N_s$ data streams in parallel to each UE. To this end, the BS employs linear precoding filters $\bm{F}_u \in \mbb{C}^{N_t \times N_s}$, $u \in \{1,\ldots,U \}$, to encode the $N_s$ data streams corresponding to UE $u$ into the $N_t$ BS antennas. Then, UE $u$ applies the combining filter $\bm{W}_u \in \mbb{C}^{N_r \times N_s}$ to the signals received from its $N_r$ antennas to estimate its corresponding $N_s$ data streams.

Assuming narrow-band block fading, the input-output relationship of our system model can be written as
\begin{equation} \label{eq:sysmodel}
\bm{y}_u = \Wu^\hermit \bm{H}_u \Fu \bm{s}_u + \sum_{\substack{j=1\\j\neq u}}^U \Wu^\hermit \bm{H}_u \bm{F}_j \bm{s}_j  + \Wu^\hermit \bm{b}_u \in \mbb{C}^{N_s},
\end{equation}
where $\bm{H}_u \in \mbb{C}^{N_r \times N_t}$ denotes the downlink channel matrix between the BS and UE $u$, $\bm{s}_u \in \mbb{C}^{N_s}$ the data symbols intended to UE $u$ and $\bm{b}_u \in \mbb{C}^{N_r}$ the noise vector. We assume that $\Rssu = \esp{\bm{s}_u\bm{s}_u^\hermit} = (1/N_s)\bm{I}_{N_s}$ and $\bm{b}_u \sim \mc{CN}(\bm{0}_{N_r \times 1}, \sigma_n^2 \bm{I}_{N_r})$ for all $u \in \{1,\ldots, U\}$. The total transmit power of the BS is denoted by $P_t$ and the system signal to noise ratio is defined as $\text{SNR} = P_t / \sigma_n^2$. It is possible to improve the achievable sum rate by optimizing the power allocation. However, we assume equal power allocation among users for analysis simplicity. The precoding matrices are thus designed to satisfy the power constraint $\frob{ \Fu }^2 = P_t/U$.

\subsection{Channel Model} \label{sec:chanmod}

We model double-sided massive MIMO channels using the narrow-band clustered channel model with $L$ paths \cite{heath2016overview,ilayach_spatially_2014,ni2016hybrid}. The downlink channel matrix $\bm{H}_u \in \mbb{C}^{N_r \times N_t}$ between the BS and UE $u$ can be expressed as
\begin{align}
&\bm{H}_u = \label{eq:channel_vector}\\ 
&\sqrt{\frac{N_t N_r}{L}}\sum_{\ell=1}^{L} \alpha_{\ell,u} \bm{a}_{r,u}\left(\phi_\ell^{(r,u)}, \theta_\ell^{(r,u)}\right) \bm{a}_{t,u}^\tran\left(\phi_\ell^{(t,u)},\theta_\ell^{(t,u)}\right)\nonumber
\end{align}
where $\alpha_{\ell,u}$ denotes the complex channel gain of path $\ell$, $\bm{a}_{t,u} \in \mbb{C}^{N_t}$ and $\bm{a}_{r,u} \in \mbb{C}^{N_r}$ the transmit and receive array response vectors evaluated at azimuth $\{\phi_\ell^{(t,u)}, \phi_\ell^{(r,u)}\}$ and elevation $\{\theta_\ell^{(t,u)},\theta_\ell^{(r,u)}\}$ angle pairs, respectively. The departure and arrival angles are taken from continuous distributions which depend on the environment. We assume that all paths are statistically independent and that the number of paths $L$ is the same for all BS-UE links to simplify the analysis. This can be achieved by selecting the $L$ strongest paths for each link. We model the complex channel gains $\alpha_{\ell,u}$ as independent and identically distributed (i.i.d.) circular symmetric Gaussian random variables with zero mean and variance $\sigma_\alpha^2$. At mmWave bands, the number of paths $L$ is typically much smaller than the numbers of antennas $N_t,\,N_r$ at BS and UE, respectively~\cite{hemadeh2017millimeter}. Using matrix notation, \eqref{eq:channel_vector} can be rewritten as 

{\small
\begin{gather}
\bm{H}_u = \bm{A}_{r,u} \bm{\Gamma}_u \bm{A}_{t,u}^\tran, \label{eq:chanmat}\\
\bm{A}_{t,u} = \left[ \bm{a}_{t,u}\left(\phi_1^{(t,u)},\theta_1^{(t,u)}\right), \ldots, \bm{a}_{t,u}\left(\phi_L^{(t,u)},\theta_L^{(t,u)}\right)  \right] \in \mbb{C}^{N_t \times L},\\
\bm{A}_{r,u} = \left[ \bm{a}_{r,u}\left(\phi_1^{(r,u)},\theta_1^{(r,u)}\right), \ldots, \bm{a}_{r,u}\left(\phi_L^{(r,u)},\theta_L^{(r,u)}\right)  \right] \in \mbb{C}^{N_r \times L},\\
\bm{\Gamma}_u = \sqrt{\frac{N_t N_r}{L}}\Diag(\alpha_{1,u},\ldots,\alpha_{L,u}) \in \mbb{C}^{L\times L}.
\end{gather}}

The rank of $\bm{H}_u$ depends on the angular distribution of the paths. For example, if the angles are independently taken from a uniform distribution and assuming $L \leq \min(N_t,N_r)$, then we have that $\rank(\bm{H}_u) = L$ with probability $1$.

In our simulations, we consider uniform linear arrays (ULAs) at both transmit and receive sides without loss of generality. In fact, any type of array geometry compatible with \eqref{eq:channel_vector} is valid for this work. The considered ULAs are comprised of omni-directional antennas with inter-antenna  spacing of $d = \lambda/2$, where $\lambda$ denotes the carrier wavelength. Therefore, the array response vectors are written as
\begin{align} \label{eq:aresp}
\bm{a}_{x,u}(\phi) = 1/\sqrt{N_x} \left[ 1,\, e^{-\jmath \pi \cos \phi}, \ldots, e^{-\jmath \pi (N_x-1)\cos \phi} \right]^\tran
\end{align}
for $x \in \{t, r\}$ and $\phi \in (-\pi,\pi)$. 

\subsection{Layered Transceiver Architecture}

We consider the layered filtering architecture proposed in~\cite{alkhateeb2017multilayer} to tackle the large dimensionality of double-sided massive MIMO systems. This filtering scheme consists of factorizing the filter matrix into outer and inner filter matrices. The former serves to form a low-dimensional \emph{effective} MIMO channel while the latter implements the precoding or combining operation. The precoding filter matrix $\Fu$ is thus decomposed into an outer factor $\Fou \in \mbb{C}^{N_t \times M_t}$ and an inner factor $\gamma_u\Fiu \in \mbb{C}^{M_t \times N_s}$ as $\Fu = \gamma_u \Fou \Fiu$, with $M_t \leq N_t$. Likewise, the combining matrix is factorized as $\Wu = \Wou \Wiu$, where $\Wou \in \mbb{C}^{N_r \times M_r}$ and $\Wiu \in \mbb{C}^{M_r \times N_s}$ with $M_r \leq N_r$. We define the normalization factor 
\begin{equation} \label{eq:norm}
\gamma_u =  \frac{\sqrt{P_t/U}}{\frob{\Fou \Fiu}}
\end{equation}
to satisfy the transmit power constraint $\frob{\Fu}^2 = P_t/U$.

Regarding the hardware implementation of the transceiver system, the multi-layer scheme can be implemented in both fully-digital and hybrid A/D strategies~\cite{caire2013jsdm,alkhateeb2017multilayer}. In the former strategy, the precoder $\Fu = \gamma_u \Fou \Fiu$ and combiner $\Wu = \Wou \Wiu$ are completely implemented in baseband. In the latter strategy, the outer filters $\Fou$ and $\Wou$ are implemented in the analog domain and the inner filters $\gamma_u\Fiu$ and $\Wiu$ are built on baseband. In the hybrid A/D strategy, the outer filters are constrained by the analog hardware with, for example, elementwise constant-modulus restriction~\cite{heath2016overview,ribeiro2018energy}. This constraint can be avoided by spending two analog phase shifters for each beamforming coefficient, as described in~\cite{lin2017thic}. Such hardware constraints are not necessary when the transceiver filters are completely implemented in baseband, as in the fully-digital architecture. Of course, the transceiver design should mind other hardware-related constraints such as total or per-antenna power constraint, peak-to-average power ratio, among others.

Let us define the effective channel matrices:
\begin{equation} \label{eq:effchan}
\bm{H}_{\text{eff},u,j} = \Wou^\hermit \bm{H}_u \Foj \in \mbb{C}^{M_r \times M_t},
\end{equation}
for all $u,\,j \in \{1,\ldots,U\}$. If $u=j$, then \eqref{eq:effchan} is simply written as  $\bm{H}_{\text{eff},u} = \Wou^\hermit \bm{H}_u \bm{F}_{\text{o},u}$. We also define the effective outer-layer-filtered noise $\bm{b}_{\text{eff},u} = \Wou^\hermit \bm{b}_u \in \mbb{C}^{M_r}$. Note that $\bm{b}_{\text{eff},u} \sim \mc{CN}(\bm{0}_{M_r\times 1}, \sigma_n^2 \Wou^\hermit \Wou)$. For future convenience, let us rewrite \eqref{eq:sysmodel} in terms of the effective channels and inner layer filters:
\begin{gather} \label{eq:effsysmodel}
\bm{y}_u = \gamma_u \Wiu^\hermit \Heffu \Fiu \bm{s}_u + \\
 \sum_{\substack{j=1\\j\neq u}}^U \gamma_j\Wiu^\hermit \Heffuj \Fij \bm{s}_j + \Wiu^\hermit \bm{b}_{\text{eff},u} \in \mbb{C}^{N_s}.
\end{gather}

\subsection{Channel State Information Acquisition} \label{sec:csi}

We assume that our double-sided massive MIMO system operates on perfectly synchronized TDD. The CSI acquisition is divided into two stages. First, the CSI necessary to compute the outer layer filters is obtained. We consider the following acquisition scenarios for outer layer CSI:
\begin{itemize}
	\item Statistical CSI -- The BS and the UE estimate  $\Cul = \esp{\bm{H}_u^\hermit \bm{H}_u} $ and $\Cdl = \esp{\bm{H}_u\bm{H}_u^\hermit} $, respectively, over some time slots. Subspace estimation~\cite{vallet2012improved} or compressive sensing-based approaches~\cite{park2018spatial} can be used to estimate the statistical CSI;
	\item Partial CSI -- Both BS and UE have perfect knowledge of the macroscopic channel parameters: the path power $|\alpha_{\ell,u}|^2$ and azimuth angles $\phi_\ell^{(t,u)}$ and $\phi_\ell^{(r,u)}$. Channel estimation methods that exploit the mmWave channel sparsity can be considered to obtain the partial CSI~\cite{alkhateeb2014channel,li2017millimeter}.
\end{itemize}
We would like to emphasize that the outer layer filters depend only on macroscopic CSI (path power and angular directions). The statistical CSI depends only on the path power and on the angles (via the antenna array response vectors) and it does not rely on microscopic channel variations (phase-shifts of the individual multipath components), which are averaged out with the statistical expectation in $\Cul$ and $\Cdl$.

The second CSI acquisition stage consists of estimating the inner layer effective channels $\Heffuj$ (inner layer CSI). The inner layer CSI acquisition task is not expensive due to the low dimensions of the effective channel matrices. It can be efficiently performed by well-known MMSE estimators~\cite{marzetta2016fundamentals} and CSI feedback methods~\cite{schwarz2019grassmannian,love2005limited,schwarz2013adaptive} without much overhead. Therefore, we consider that both BS and UE have perfect knowledge of the effective channels for analysis simplicity. Assessing the impact of imperfect CSI on the proposed transceiver strategies is out of the scope of this work. The inner layer filters depend on microscopic channel variations, which change quickly with movements in the order of the wavelength and cause microscopic fading.

The outer and inner layer filters are updated according to the different time scales. The macroscopic CSI necessary for the outer layer filters does not change significantly as long as the receiver stays within the $3$-dB beamwidth of the transmitter's antenna array. If the distance between the transmitter and the receiver is at least several tens of meters, then the receiver will be within the $3$-dB beamwidth for some time provided that it is not moving too fast. By contrast, the microscopic CSI changes faster, even than the channel coherence time. In conclusion, the inner layer filters are updated more often than the outer layer filters because of the different time scales of the corresponding CSI.

\section{Transceiver Schemes} \label{sec:trx}

We present low-complexity outer and inner layer filtering methods for double-sided massive MIMO systems in this section. The filtering layers are designed to perform different tasks: the outer layer typically aims to provide an SNR gain, whereas the inner layer seeks to cancel multi-user interference~\cite{alkhateeb2017multilayer}. We study three outer layer schemes, namely 
\begin{itemize}
	\item Covariance matrix eigenfilter (CME);
	\item Power-dominant path selection (PPS) method;
	\item Semi-orthogonal path selection (SPS) method
\end{itemize}
and four methods for the inner filtering layer:
\begin{itemize}
	\item Maximum eigenmode transmission (MET) and maximum eigenmode reception (MER): MET-MER;
	\item Maximum eigenmode transmission (MET) and block diagonalization (BD) reception: MET-BD;
	\item Maximum eigenmode transmission (MET) and minimum mean square error (MMSE) reception: MET-MMSE;
	\item Block diagonalization (BD) transmission and maximum eigenmode reception (MER): BD-MER.
\end{itemize}
It is desirable to form full-rank effective channels $\Heffu$ so that the proposed transceiver schemes support multi-stream transmission. Therefore, we consider the following assumptions:
\begin{enumerate}
	\item[A1] The rank of the channel matrices is lower bounded as
	\[
	\min(M_r,M_t) \leq \rank(\bm{H}_u) = L
	\] for all $u \in \{1,\ldots,U\}$;
	\item[A2] The outer layer filters have full rank, i.e., $\rank(\Wou) = M_r$ and $\rank(\Fou) = M_t$. 
\end{enumerate}
We have that $\rank(\Heffu) = \min(M_r,M_t)$ as a consequence of A1 and A2. A1 is satisfied provided that the channel has enough degrees of freedom, which depends on the assumed channel properties. Finally, A2 can be enforced when designing the outer layer filters, as we will show in the following.

\subsection{Outer Layer Filtering} \label{sec:outer}

\subsubsection{Covariance Matrix Eigenfilter (CME)} \label{sec:cme}

Assuming statistical CSI, let
\begin{gather}
\hat{\bm{C}}_{\text{dl},u} = \bm{Q}_{\text{dl},u} \bm{\Xi}_{\text{dl},u} \bm{Q}_{\text{dl},u}^\hermit, \label{eq:cdl} \\
\hat{\bm{C}}_{\text{ul},u} = \bm{Q}_{\text{ul},u} \bm{\Xi}_{\text{ul},u} \bm{Q}_{\text{ul},u}^\hermit \label{eq:cul}
\end{gather}
denote the eigendecomposition of the estimated channel covariance matrices,  $\bm{Q}_{\text{dl},u} \in \mbb{C}^{N_r \times N_r}$, $\bm{Q}_{\text{ul},u} \in \mbb{C}^{N_t \times N_t}$ the eigenvector matrices, and $\bm{\Xi}_{\text{dl},u} \in \mbb{C}^{N_r \times N_r}$, $\bm{\Xi}_{\text{ul},u} \in \mbb{C}^{N_t \times N_t}$ the corresponding  eigenvalue matrices. The outer layer filters $\Wou$ and $\Fou$ are derived as the $M_r$ and $M_t$ dominant eigenvectors of $\hat{\bm{C}}_{\text{dl},u} $ and $\hat{\bm{C}}_{\text{ul},u}$, respectively. Define $\tilde{\bm{Q}}_{\text{dl},u} $ and $\tilde{\bm{Q}}_{\text{ul},u} $ as the truncated eigenvector matrices with the $M_r$ and $M_t$ first columns of the corresponding matrices. Then, the eigenfilters are given by~\cite{alkhateeb2017multilayer}
\begin{equation}
\Fou = \tilde{\bm{Q}}_{\text{ul},u} \in \mbb{C}^{N_t \times M_t}, \quad \Wou = \tilde{\bm{Q}}_{\text{dl},u} \in \mbb{C}^{N_r \times M_r}
\end{equation}
for all $u \in \{1,\ldots, U\}$. We hereafter refer to this filtering scheme as covariance matrix eigenfilter (CME).

\subsubsection{Power-dominant Path Selection (PPS)}

Considering partial CSI, the power-dominant path selection (PPS) naively selects the $M_t$ and $M_r$ dominant paths to form the outer layer filters. Let $\mc{L}_D^{(t)}$ and $\mc{L}_D^{(r)}$ denote sets containing the indices of the $M_t$ and $M_r$ dominant paths, respectively. Then 
\begin{equation}
\Fou = [ \bm{a}_{t,u}(\phi_{\ell_t}^{(t,u)})], \quad \Wou = [ \bm{a}_{r,u}(\phi_{\ell_r}^{(r,u)}) ]
\end{equation}
for all $\ell_t \in \mc{L}_D^{(t)}$ and $\ell_r \in \mc{L}_D^{(r)}$.

\subsubsection{Semi-orthogonal Path Selection (SPS)}

Although the PPS method is simple, it has a major drawback: it may select highly correlated paths, which would yield rank-deficient effective channels. That would not be ideal for a multi-stream communications scenario. As an alternative to SPS and CME, we propose a novel sub-optimal solution which selects the beamforming directions using a semi-orthogonal path selection (SPS) algorithm. The proposed solution can be seen as a customization of the semi-orthogonal user selection algorithm of~\cite{yoo2006optimality} to the beamforming problem. SPS is presented in Algorithm~\ref{alg:sps} considering
\begin{itemize}
	\item a general array manifold matrix $\bm{A} = \left[ \bm{a}_\ell \right] \in \mbb{C}^{N \times L}$;
	\item a path power vector $[|\alpha_1|^2,\ldots,|\alpha_L|^2]^\tran$;
	\item $M \leq L$ desired paths.
\end{itemize}
Partial CSI knowledge is sufficient here, since the array manifold matrix $\bm{A}$ can be built from the departure or arrival angles in partial CSI, as in \eqref{eq:aresp}.

SPS seeks $M$ semi-orthogonal steering vectors with relatively strong power. Semi-orthogonality is enforced by steps $2$ and $4$ in Algorithm~\ref{alg:sps}: the non-selected path components in $\Lambda_i$ are projected onto the orthogonal complement of $\vecspan\left[ \bm{g}_{(1)},\ldots,\bm{g}_{(i-1)} \right]$. Then, among these semi-orthogonal vectors, the path with largest power, measured by $\neuclid{\bm{g}_\ell}^2$ is selected in Step $3$. Since SPS provides outer layer precoding and combining matrices formed by $M_t$ and $M_r$ columns of $\bm{A}_{t,u}$ and $\bm{A}_{r,u}$, respectively, then it can be shown that $\frob{\Fou}^2 = M_t$ and $\frob{\Wou}^2=M_r$. In summary, the outer layer filters for the BS-UE link $u$ are chosen as
\begin{enumerate}
	\item $\Fou \gets \text{SPS}(\bm{A}_{t,u},[|\alpha_{1,u}|^2,\ldots,|\alpha_{L,u}|^2]^\tran,\, M_t)$;
	\item $\Wou \gets \text{SPS}(\bm{A}_{r,u},[|\alpha_{1,u}|^2,\ldots,|\alpha_{L,u}|^2]^\tran,\, M_r)$.
\end{enumerate}

\begin{algorithm}
	\small
	\caption{Semi-orthogonal Path Selection (SPS)}
	\label{alg:sps}
	\begin{algorithmic}[1]
		\Procedure{SPS}{$\bm{A}$, $[|\alpha_{1}|^2,\ldots,|\alpha_{L}|^2]^\tran$, $M$}
		\State \underline{Step 1: Initialization:}
		\State $\Lambda_1 \gets \{1,\ldots,L\}$ \Comment{Non-selected paths set}
		\State $S \gets \text{Empty set}$ \Comment{Selected paths set}
		\State $i \gets 1$
		\While{$\#(S) < M$} \label{sps:loop1}
		\State \underline{Step 2: Form orthogonal projections:}
		\For{each path $\ell \in \Lambda_i$} \label{sps:loop2}
		\State $\bm{g}_{\ell} \gets |\alpha_\ell|^2\bm{a}_{\ell}$ 
		\If{$i\geq 2$}
			\State $\bm{g}_\ell \gets |\alpha_\ell|^2\bm{a}_\ell - \sum_{j=1}^{i-1} \bm{g}_{(j)} \frac{\bm{g}_{(j)}^\hermit(|\alpha_\ell|^2\bm{a}_\ell) }{\neuclid{\bm{g}_{(j)}}^2} $ \label{sps:loop3}
		\EndIf 
		\EndFor 
		\State \underline{Step 3: Select $i$th path:}
		\State $\pi(i) \gets \argmax_{\ell \in \Lambda_i} \neuclid{\bm{g}_\ell}^2$
		\State $S \gets S\cup \{ \pi(i) \}$
		\State $\bm{a}_{(i)} \gets \bm{a}_{\pi(i)}$
		\State $\bm{g}_{(i)} \gets \bm{g}_{\pi(i)}$
		\State \underline{Step 4: Update non-selected paths set:}
		\State $\Lambda_{i+1} \gets \{ \ell \in \Lambda_i \mid \ell \neq \pi(i) \}$ 
		\State $i \gets i + 1$ 
		\EndWhile
		\State \Return $\bm{A}_S = [ \bm{a}_s ]$, $s \in S$.
		\EndProcedure
	\end{algorithmic}
\end{algorithm}

\subsection{Inner Layer Filtering} \label{sec:inner}

The low-dimensional effective channels $\Heffu$ can be formed once the outer layer filters have been selected.  The design of inner layer filters is now regarded as a classical multi-user MIMO transceiver design problem. For future convenience, let the singular value decomposition (SVD) of $\Heffu$ be written as
\begin{equation}
\Heffu = \left[ \bm{U}_u^s,\, \bm{U}_u^o \right] \Blkdiag\left( \bm{\Sigma}_u^{s},\,\bm{\Sigma}_u^{o} \right) \left[ \bm{V}_u^{s},\,\bm{V}_u^{o} \right]^\hermit,
\end{equation}
where $\bm{U}_u^s\in \mbb{C}^{M_r \times N_s}$ contains the $N_s$ first left singular vectors, $\bm{V}_u^{s} \in \mbb{C}^{M_t \times N_s}$ the first $N_s$ right singular vectors, $\bm{\Sigma}_u^{s} = \Diag(\sigma_1,\ldots,\sigma_{N_s})$ the matrix formed by the $N_s$ first singular values and $\bm{\Sigma}_u^{o} = \Diag(\sigma_{N_s+1},\ldots,\sigma_{\min (M_r,M_t)})$ the matrix with the remaining singular values. Note that the truncated singular vector matrices are semi-unitary, i.e., $\bm{U}_u^{s\hermit}\bm{U}_u^s = \bm{V}_u^{s\hermit}\bm{V}_u^s = \bm{I}_{N_s}$.

Regarding CSI, we make the following assumptions:
\begin{itemize}
	\item BS, as well as UEs, have perfect knowledge of the corresponding $\Heffu$ for all inner layer transceiver strategies. This is a practical assumption, since $M_t, M_r \leq N_t, N_r$, allowing the development of efficient CSI feedback methods~\cite{schwarz2019grassmannian,love2005limited,schwarz2013adaptive};
	\item MET-BD, BD-MER, MET-MMSE additionally have perfect knowledge of the interfering effective channel matrices $\Heffuj$ for all $j \neq u$ at the user-side.
\end{itemize}

\subsubsection{MET-MER: Maximum Eigenmode Transmission (MET) and Maximum Eigenmode Reception (MER)}

The MET-MER transceiver scheme selects the inner precoding matrix $\Fiu$ as the first $N_s$ right singular vectors of $\Heffu$ and the inner combining matrix $\Wiu$ as the first $N_s$ left singular vectors of $\Heffu$~\cite{schwarz2014exploring}: 
\begin{gather} \label{eq:metmertrx}
	\Fiu = \bm{V}_u^s \in \mbb{C}^{M_t \times N_s},\quad\Wiu = \bm{U}_u^{s} \in \mbb{C}^{M_r \times N_s}.
\end{gather}
The MET-MER transceiver seeks to maximize the SNR at the UE disregarding multi-user interference. The BS can transmit up to $N_s \leq \min(M_r,M_t)$ data streams per user simultaneously. 

\subsubsection{MET-BD: Maximum Eigenmode Transmission (MET) and Block Diagonalization (BD) Reception}

In this scheme, the UE satisfies the BD condition to cancel multi-user interference~\cite{ni2016hybrid}:
\begin{gather}
\rank\left(\Wiu^\hermit \Heffu\right) \stackrel{!}{=} N_s,\quad u \in \{1,\ldots,U\}, \label{eq:metbdcond1}\\
\Wiu^\hermit \bar{\bm{H}}_{\text{eff},u} \stackrel{!}{=} \bm{0}_{N_s \times (U-1)N_s},\quad u \in \{1,\ldots,U\}, \label{eq:metbdcond2}\\
\bar{\bm{H}}_{\text{eff},u} =  \left[ \bm{H}_{\text{eff},u,1}\bm{F}_{\text{i},1}, \ldots, \bm{H}_{\text{eff},u,u-1}\bm{F}_{\text{i},u-1}, \right. \\
\left. \bm{H}_{\text{eff},u,u+1}\bm{F}_{\text{i},u+1},\ldots,\bm{H}_{\text{eff},u,U}\bm{F}_{\text{i},U} \right]  \in\mbb{C}^{M_r \times (U-1)N_s}. \label{eq:metbdmat}
\end{gather}
where $\Heffuj$ is defined in \eqref{eq:effchan}, and $\bm{F}_{\text{i},j} = \bm{V}_j^s$, for all $j \in \{1,\ldots,U\} \setminus \{u\}$. The BD combiner requires $U N_s \leq M_r$ in order to simultaneously cancel the multi-user interference and  allow the transmission of $N_s$ data streams per user. If this condition is satisfied, then $(U-1)N_s \leq M_r$ and $\bar{\bm{H}}_{\text{eff},u}$ becomes full column rank. Consequently, interfering users can be canceled by projecting $\Wiu$ onto the null-space of $\bar{\bm{H}}_{\text{eff},u}^\hermit$. We project the MER combiner~\eqref{eq:metmertrx} onto the null-space of the multi-user interference matrix $\bar{\bm{H}}_{\text{eff},u}^\hermit$ to maximize the intended signal power while canceling interference. Let the SVD of $\bar{\bm{H}}_{\text{eff},u}$ be 
\begin{equation} \label{eq:mui_m_metbd}
	\bar{\bm{H}}_{\text{eff},u} = \left[ \bar{\bm{U}}_u^s,\, \bar{\bm{U}}_u^o \right] \Blkdiag\left( \bar{\bm{\Sigma}}_u^{s},\,\bar{\bm{\Sigma}}_u^{o} \right) \left[ \bar{\bm{V}}_u^{s},\,\bar{\bm{V}}_u^{o} \right]^\hermit,
\end{equation}
where $\bar{\bm{U}}_u^o \in \mbb{C}^{M_r \times (U-1)N_s}$ contains the last $(U-1)N_s$ left singular vectors of $\bar{\bm{H}}_{\text{eff},u}$. The null-space projection matrix is defined as $\bar{\bm{P}} = \bar{\bm{U}}_u^o \bar{\bm{U}}_u^{o\hermit} \in \mbb{C}^{M_r \times M_r}$. The MET-BD transceiver filters are thus given by:
\begin{equation}
	\Fiu = \bm{V}_u^{s} \in \mbb{C}^{M_t \times N_s},\quad \Wiu = \bar{\bm{P}} \bm{U}_{u}^s \in \mbb{C}^{M_r \times N_s}.
\end{equation} 

\subsubsection{MET-MMSE: Maximum Eigenmode Transmission (MET) and Minimum Mean Square Error (MMSE) Reception}

We also consider interference-aware MMSE combining~\cite{schwarz2013antenna} to balance between the multi-user interference minimization and intended user power maximization. The MMSE inner layer filter is obtained from
\begin{equation} \label{eq:mmse}
\min_{\Wiu \in \mbb{C}^{M_r \times N_s}} \,\,\esp{ \neuclid{\bm{s}_u - \bm{y}_u}^2 },
\end{equation}
where $\bm{y}_u$ is the received signal at UE $u$ defined in \eqref{eq:effsysmodel} and the expectation is performed with respect to the transmitted symbols and additive noise. By solving \eqref{eq:mmse} and setting the MET precoders $\Fiu = \bm{V}_u^s$ for all $u \in \{1,\ldots,U\} $, the MMSE combiner reads as~\cite{schwarz2013antenna}:

{\small\begin{gather}
\Wiu = \frac{\gamma_u}{N_s} \bm{R}_{yy}^{-1}  \Heffu \Fiu, \label{eq:mmsef1}\\
\bm{R}_{yy} = \sigma_n^2 \Wou^\hermit\Wou + \sum_{j = 1}^U \frac{|\gamma_j|^2}{N_s} \Heffuj \Fij \Fij^\hermit \Heffuj^\hermit. \label{eq:mmsef2}
\end{gather}}
Note that the MMSE combiner does not require $UN_s \leq M_r$ unlike the BD combiner.

\subsubsection{BD-MER: Block Diagonalization (BD) Transmission and Maximum Eigenmode Reception (MER)}

With this strategy, the block diagonalization condition is formulated at the transmitting side~\cite{ni2016hybrid}:
\begin{gather}
\rank\left(\Heffu \Fiu \right) \stackrel{!}{=} N_s,\quad u \in \{1,\ldots,U \}, \label{eq:bdmercond1}\\
\tilde{\bm{H}}_{\text{eff},u} \Fiu \stackrel{!}{=} \bm{0}_{(U-1)N_s \times N_s},\quad u\in \{1,\ldots,U\}, \label{eq:bdmercond2}\\
\tilde{\bm{H}}_{\text{eff},u} = \left[ (\bm{W}_{\text{i},1}^\hermit\bm{H}_{\text{eff},1,u})^\hermit,\ldots,
												      (\bm{W}_{\text{i},u-1}^\hermit\bm{H}_{\text{eff},u-1,u})^\hermit, \ldots, \right.\\
												      \left.
												      (\bm{W}_{\text{i},U}^\hermit\bm{H}_{\text{eff},U,u})^\hermit\right]^\hermit \in \mbb{C}^{(U-1)N_s \times M_t}. \label{eq:bdmerint}
\end{gather}
with $\bm{W}_{\text{i},j} = \bm{U}_j^s$, for all $j \in \{1,\ldots,U\} \setminus \{u\}$. The BD precoder is able to mitigate multi-user interference at the BS and transmit the $N_s$ data streams per user when $UN_s \leq M_t$. In this case, $\tilde{\bm{H}}_{\text{eff},u}$ is of full row rank and the precoding filter lies in the null-space of $\tilde{\bm{H}}_{\text{eff},u}$. We project the MET precoder~\eqref{eq:metmertrx} onto the null-space of the multi-user interference matrix $\tilde{\bm{H}}_{\text{eff},u}$ to maximize the power of the intended UE while mitigating interference at non-intended UEs. Let the SVD of $\tilde{\bm{H}}_{\text{eff},u}$ be 
\begin{equation} \label{eq:mui_m_bdmer}
	\tilde{\bm{H}}_{\text{eff},u} = \left[ \tilde{\bm{U}}_u^s,\, \tilde{\bm{U}}_u^o \right] \Blkdiag\left( \tilde{\bm{\Sigma}}_u^{s},\,\tilde{\bm{\Sigma}}_u^{o} \right) \left[ \tilde{\bm{V}}_u^{s},\,\tilde{\bm{V}}_u^{o} \right]^\hermit,
\end{equation}
where $\tilde{\bm{V}}_u^o \in \mbb{C}^{M_t \times (U-1)N_s}$ contains the last $(U-1)N_s$ right singular vectors of $\tilde{\bm{H}}_{\text{eff},u}$. The null-space projection matrix is written as $\tilde{\bm{P}} = \tilde{\bm{V}}_u^o\tilde{\bm{V}}_u^{o\hermit} \in \mbb{C}^{M_t \times M_t}$. Therefore, the BD-MER transceiver filters are given by:
\begin{equation}
	\Fiu = \tilde{\bm{P}} \bm{V}_u^s \in \mbb{C}^{M_t \times N_s},\quad \Wiu = \bm{U}_u^s \in \mbb{C}^{M_r \times N_s}.
\end{equation}

\subsection{Complexity Analysis} \label{sec:comp}

In this section, we evaluate the complexity of the proposed transceiver strategies. The total complexity is divided into three parts: outer layer filter design, effective channel matrices computation and inner layer filter design.

\paragraph{Outer Layer Filter Design}

\begin{itemize}
	\item CME -- The computational complexity of eigendecompositions \eqref{eq:cdl} and \eqref{eq:cul} is $O(N_r^3)$ and $O(N_t^3)$, respectively;
	\item PPS -- This method involves sorting an $L$-dimensional vector. This operation can be carried out with complexity $O(L \log L)$;
	\item SPS -- This algorithm has complexity $O(LM^2 N^2)$, where $M \in \{ M_t,\,M_r \}$ and $N \in \{N_t,\,N_r\}$. The factor $N^2$ refers to orthogonal projections and $LM^2$ to the loops in lines~\ref{sps:loop1},~\ref{sps:loop2} and~\ref{sps:loop3} (Algorithm~\ref{alg:sps}).
\end{itemize}

\paragraph{Effective Channel Matrices} The computation of all effective matrices~\eqref{eq:effchan}, $\forall u,\,j \in \{1,\ldots,U\}$, has complexity $O[ (N_r N_t M_t + M_r M_t N_r) U^2]$. The factor $(N_r N_t M_t + M_r M_t N_r)$ refers to the calculation of a single $\bm{H}_{\text{eff},u,j}$ matrix and $U^2$ to the computation of all $(u,j)$ combinations.

\paragraph{Inner Layer Filter Design}

\begin{itemize}
	\item MET-MER -- The eigendecompositions of $\Heffu^\hermit \Heffu$ (MET) and $\Heffu \Heffu^\hermit$ (MER) have complexity $O(M_t^3)$ and $O(M_r^3)$, respectively;
	\item MET-BD -- Forming $\bar{\bm{H}}_{\text{eff},u}$ for all UEs has complexity $O
	[M_t M_r N_s U(U-1)]$. For the MET precoder, the eigendecomposition of $\Heffu^\hermit \Heffu$ has complexity $O(M_t^3)$. For the BD combiner, the eigendecompositions of $\Heffu \Heffu^\hermit$ (MER) and $\bar{\bm{H}}_{\text{eff},u}\bar{\bm{H}}_{\text{eff},u}^\hermit$ (null-space projection matrix calculation) have complexity $O(M_r^3)$;
	\item MET-MMSE -- The MET precoder has complexity $O(M_t^3)$. For the MMSE combiner, \eqref{eq:mmsef1} and \eqref{eq:mmsef2} have complexity $O(N_s M_r M_t + N_s M_r^2)$ and $O[N_rM_r^2 + U(2M_r M_t N_s + M_r^2 N_s)]$, respectively. Note that the calculation of $\bm{R}_{yy}^{-1}$ requires  $O(M_r^3)$ operations;
	\item BD-MER -- Forming $\tilde{\bm{H}}_{\text{eff},u}$ for all UEs has complexity $O
	[M_t M_r N_s U(U-1)]$. For the BD precoder, the eigendecompositions of $\Heffu^\hermit \Heffu$ (MET) and $\tilde{\bm{H}}_{\text{eff},u}^\hermit \tilde{\bm{H}}_{\text{eff},u}$ (null-space projection matrix calculation) have complexity $O(M_t^3)$. For the MER combiner, the eigendecomposition of $\Heffu \Heffu^\hermit$ has complexity $O(M_r^3)$.
\end{itemize}

In our multi-layer approach, the outer layer filters are updated once the macroscopic CSI is outdated, whereas the inner layer filters are recalculated as the microscopic CSI changes. Fortunately, the macroscopic CSI evolves slower than the microscopic CSI, as discussed in Section~\ref{sec:csi}, therefore, the outer layer is updated once in a while, whereas the inner layer is updated more often. If $M_t$ and $M_r$ are much smaller than $N_t$ and $N_r$, then the proposed solution is less complex than the classical single-layer approach, which consists of applying the inner layer schemes directly to the $(N_r \times N_t)$-dimensional channel matrices $\bm{H}_u$. In this case, the complexity of each transceiver would be cubic with $N_t$ and $N_r$, instead of $M_t$ and $M_r$, as we observe in the proposed multi-layer approach. Moreover, the single-layer transceiver filters would be updated at the microscopic CSI timescale. The double-sided massive MIMO transceiver schemes proposed in~\cite{buzzi2018efficiency} would face similar computational challenges as the single-layer approach, because they work directly with $(N_r \times N_t)$-dimensional channel matrices.

\section{Simulation Results} \label{sec:sim}

In this section, we present and discuss a variety of numerical simulations conducted to investigate the proposed double-sided massive MIMO transceiver architectures. We are mostly interested in evaluating the spatial multiplexing capabilities of the proposed methods and identifying the most suited strategy for different channel propagation scenarios. Therefore, we consider the achievable sum rate 
\begin{gather}
R = \sum_{u=1}^U \log_2 \det \left( \bm{I}_{N_s} + \bm{C}_{u}^{-1} \bm{R}_u \right), \label{eq:rate}\\
\bm{C}_u = \sigma_n^2 \bm{W}_u^\hermit\bm{W}_u + \sum_{\substack{j=1\\j\neq u}}^U \frac{1}{N_s}\bm{W}_u^\hermit \bm{H}_u \bm{F}_j \bm{F}_j^\hermit \bm{H}_u^\hermit \bm{W}_u,\\
\bm{R}_u = \frac{1}{N_s}\bm{W}_u^\hermit \bm{H}_u \bm{F}_u \bm{F}_u^\hermit \bm{H}_u^\hermit \bm{W}_u,
\end{gather}
as the figure of merit. In our simulations, we generate the arrival and departure angles in \eqref{eq:chanmat} as follows: the $L$ rays are grouped in clusters of $4$ rays. For each cluster, we select the mean cluster angle $\bar{\phi}_c$, a random variable in $\mc{U}(0^\circ,180^\circ)$, and then the angle of each ray in the cluster is modeled as a Gaussian random variable with mean $\bar{\phi_c}$ and standard deviation of $\sigma_c$ degrees.

To achieve satisfactory spatial multiplexing, the channel has to offer sufficient degrees of freedom. MmWave channels, however, are characterized by a reduced number of scatterers~\cite{akdeniz2014millimeter}, which may decrease the channel degrees of freedom. To account for these propagation differences in the spatial multiplexing performance, we study three scattering scenarios:
\begin{itemize}
	\item Poor scattering --  $2$ clusters, $L=8$ rays;
	\item Fair scattering -- $8$  clusters, $L=32$ rays;
	\item Rich scattering -- $16$ clusters, $L=64$ rays.
\end{itemize}
The ``poor'' scenario can be seen as the pessimistic setup, which can be realistic for indoor mmWave systems. The ``rich'' scenario is regarded as the optimistic case, which can be feasible for sub-$6$ GHz systems. The ``fair'' scenario plays a compromise between the pessimistic and optimistic setups. 

\begin{figure}[t]
	\centering
%
%
\definecolor{mycolor1}{rgb}{0.00000,0.44700,0.74100}%
\definecolor{mycolor2}{rgb}{0.85000,0.32500,0.09800}%
\definecolor{mycolor3}{rgb}{0.92900,0.69400,0.12500}%
\begin{tikzpicture}

\begin{axis}[%
width=2.5in,
height=2.5in,
scale only axis,
xmin=0.1,
xmax=1,
ymin=10,
ymax=55,
xlabel = {Number of streams scaling $N_s/L$},
ylabel = {Achievable sum rate [bit/s/Hz]},
axis background/.style={fill=white},
xmajorgrids,
ymajorgrids,
legend style={legend cell align=left, align=left, draw=white!15!black},
legend pos=north west
]
\addplot [color=mycolor1, line width = 1pt, mark =  asterisk, mark size = 2pt]
  table[row sep=crcr]{%
0.125	12.9044479819405\\
0.25	24.753596212119\\
0.375	33.2692444192129\\
0.5	40.7602725267052\\
0.625	45.9018171817182\\
0.75	50.7550824463138\\
0.875	52.5260328785381\\
1	52.1204986982182\\
};

\addplot [color=mycolor3, line width = 1pt, mark = triangle, mark size = 2pt]
  table[row sep=crcr]{%
0.125	16.9399004941228\\
0.25	27.5715356367788\\
0.375	35.7766438311763\\
0.5	42.2801098014025\\
0.625	46.4924645235455\\
0.75	49.8439914606293\\
0.875	51.1073928049896\\
1	48.4633574751733\\
};

\addplot [color=black, line width = 1pt, mark = o, mark size = 2pt]
  table[row sep=crcr]{%
0.125	16.9399004941228\\
0.25	27.7853688899313\\
0.375	35.9783688106115\\
0.5	41.9963938757121\\
0.625	45.7293390618085\\
0.75	47.8489060052268\\
0.875	49.0967832784212\\
1	48.4633568662778\\
};

\node(sps) at (axis cs:0.82, 50){\footnotesize SPS};
\node(cme) at (axis cs:0.95, 54){\footnotesize CME};
\node(pps) at (axis cs:0.75, 45){\footnotesize PPS};

\end{axis}
\end{tikzpicture}%
	\caption{Outer layer methods at poor scattering ($L=8$ paths).}
	\label{fig:outer_poor}
\end{figure}
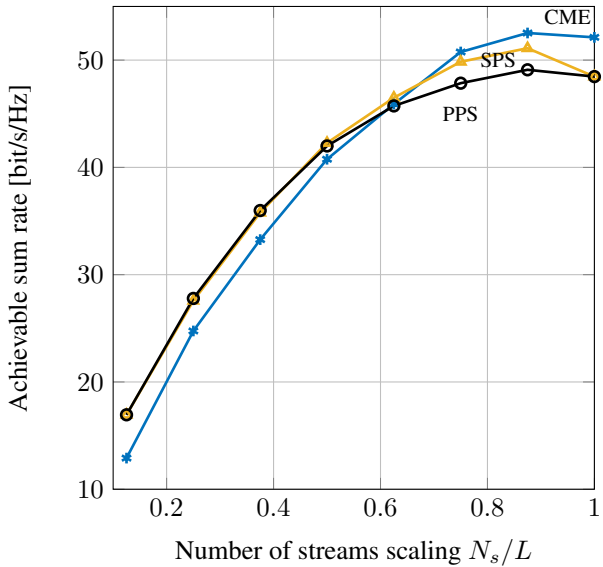

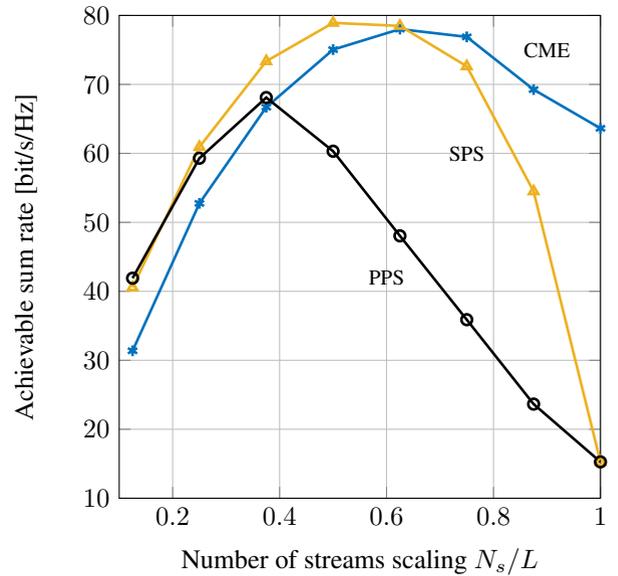
\begin{figure}[t]
	\centering
%
%
\definecolor{mycolor1}{rgb}{0.00000,0.44700,0.74100}%
\definecolor{mycolor2}{rgb}{0.85000,0.32500,0.09800}%
\definecolor{mycolor3}{rgb}{0.92900,0.69400,0.12500}%
\begin{tikzpicture}

\begin{axis}[%
width=2.5in,
height=2.5in,
scale only axis,
xmin=0.1,
xmax=1,
ymin=10,
ymax=80,
xlabel = {Number of streams scaling $N_s/L$},
ylabel = {Achievable sum rate [bit/s/Hz]},
axis background/.style={fill=white},
xmajorgrids,
ymajorgrids,
legend style={legend cell align=left, align=left, draw=white!15!black},
legend pos=south west
]
\addplot [color=mycolor1, line width = 1pt, mark =  asterisk, mark size = 2pt]
  table[row sep=crcr]{%
0.125	31.3803884764676\\
0.25	52.7545059532449\\
0.375	66.70031136838\\
0.5	75.0435005094183\\
0.625	78.0160550992513\\
0.75	76.8907658163411\\
0.875	69.2478917988629\\
1	63.6325164888812\\
};

\addplot [color=mycolor3, line width = 1pt, mark = triangle, mark size = 2pt]
  table[row sep=crcr]{%
0.125	40.5619173411145\\
0.25	60.9255894588417\\
0.375	73.3589488117284\\
0.5	78.9234241179079\\
0.625	78.4826654096723\\
0.75	72.6262987537142\\
0.875	54.4929815818471\\
1	15.2718433283381\\
};

\addplot [color=black, line width = 1pt, mark = o, mark size = 2pt]
  table[row sep=crcr]{%
0.125	41.9026052540381\\
0.25	59.3009546404763\\
0.375	68.0905613966993\\
0.5	60.3149858285961\\
0.625	48.0441317408033\\
0.75	35.8892650746243\\
0.875	23.6500969054222\\
1	15.2718429626623\\
};

\node(sps) at (axis cs:0.75, 60){\footnotesize SPS};
\node(cme) at (axis cs:0.90, 75){\footnotesize CME};
\node(pps) at (axis cs:0.6, 42){\footnotesize PPS};

\end{axis}
\end{tikzpicture}%
	\caption{Outer layer methods at fair scattering ($L=32$ paths).}
	\label{fig:outer_fair}
\end{figure}

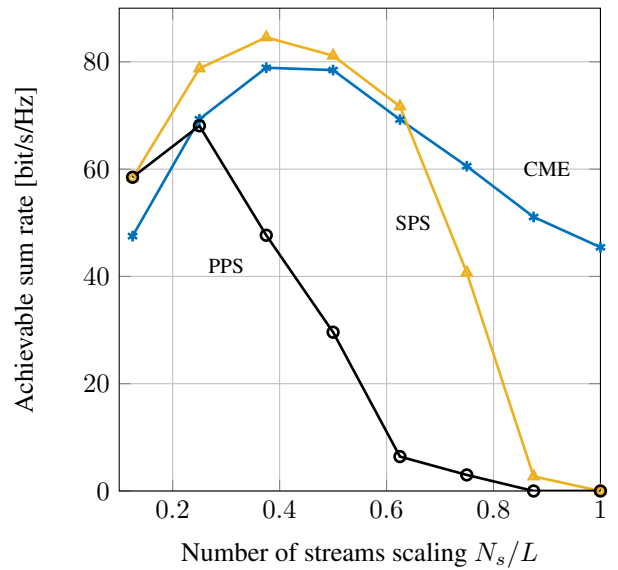
\begin{figure}[t]
	\centering
%
%
\definecolor{mycolor1}{rgb}{0.00000,0.44700,0.74100}%
\definecolor{mycolor2}{rgb}{0.85000,0.32500,0.09800}%
\definecolor{mycolor3}{rgb}{0.92900,0.69400,0.12500}%
\begin{tikzpicture}

\begin{axis}[%
width=2.5in,
height=2.5in,
scale only axis,
xmin=0.1,
xmax=1,
ymin=0,
ymax=90,
xlabel = {Number of streams scaling $N_s/L$},
ylabel = {Achievable sum rate [bit/s/Hz]},
axis background/.style={fill=white},
xmajorgrids,
ymajorgrids,
legend style={legend cell align=left, align=left, draw=white!15!black}
]
\addplot [color=mycolor1, line width = 1pt, mark =  asterisk, mark size = 2pt]
  table[row sep=crcr]{%
0.125	47.5085239486021\\
0.25	69.3040858517676\\
0.375	78.8988106123378\\
0.5	78.4602398323448\\
0.625	69.2408296763791\\
0.75	60.551203137559\\
0.875	51.076477590326\\
1	45.4381962270351\\
};

\addplot [color=mycolor3, line width = 1pt, mark = triangle, mark size = 2pt]
  table[row sep=crcr]{%
0.125	58.3679305462209\\
0.25	78.7651711924228\\
0.375	84.5631861556981\\
0.5	81.1485979129183\\
0.625	71.6840417834081\\
0.75	40.7197617666973\\
0.875	2.71686624056586\\
1	0\\
};

\addplot [color=black, line width = 1pt, mark = o, mark size = 2pt]
  table[row sep=crcr]{%
0.125	58.4883503130138\\
0.25	68.0731592755454\\
0.375	47.6755779722828\\
0.5	29.5973214720336\\
0.625	6.40945659480009\\
0.75	2.99627111979947\\
0.875	0\\
1	0\\
};

\node(sps) at (axis cs:0.65, 50){\footnotesize SPS};
\node(cme) at (axis cs:0.90, 60){\footnotesize CME};
\node(pps) at (axis cs:0.3, 42){\footnotesize PPS};

\end{axis}
\end{tikzpicture}%
	\caption{Outer layer methods at rich scattering ($L=64$ paths).}
	\label{fig:outer_rich}
\end{figure}
 
We present three groups of simulation results. In the first group, we examine the outer layer filtering strategies. In the second group, we compare the achievable sum rate performance of the proposed inner layer filtering methods. In the final simulation group, we benchmark the proposed transceivers. In all simulations, we considered the following parameter setup: $N_t=N_r=64$ antennas, noise variance $\sigma_n^2 = 10^{-3}$, i.i.d. channel gains variance $\sigma_\alpha^2 = 1$ and Gaussian spreading standard deviation $\sigma_c = 5^\circ$.  The downlink and uplink channel covariance matrices for statistical CSI (Section~\ref{sec:csi}) were estimated by averaging over $100$ time slots. The presented results were averaged over $1000$ independent experiments.

\subsection{Outer Layer Filters}
Let us first compare the spatial multiplexing performance of the outer layer filtering methods. Since this layer mainly concentrates at SNR gain, we disregard multi-user interference by setting $U=1$. Furthermore, we do not employ inner layer filtering, and, thus, $\bm{F}_u$ and $\bm{W}_u$ in \eqref{eq:rate} are given by the outer layer filters with $M_t = M_r = N_s$. Let us assess the impact of the number of multiplexed data streams $N_s$. To this end, we consider the ratio $N_s/L$. The transceiver operates at maximum spatial multiplexing when $N_s/L = 1$. We set $\text{SNR} = 20\,\text{dB}$ for the results presented in figures~\ref{fig:outer_poor}--\ref{fig:outer_rich}.

In Figure~\ref{fig:outer_poor}, we evaluate the outer layer schemes at the poor scattering scenario. We observe that all methods perform roughly the same. At aggressive spatial multiplexing ($N_s/L$~approx.~$1$), CME exhibits an advantage over the geometrical methods. Since we only have a few paths in this poor setup, it is expected that SPS and PPS do not differ much. With only $2$ clusters, at least two paths will likely show some spatial correlation. Figure~\ref{fig:outer_fair} reveals that PPS tends to perform worse as we increase the number of paths. This is because of the likelihood of the strongest paths being spatially correlated increases with $L$. Moreover, we observe that SPS outperforms PPS because it avoids selecting highly correlated paths, which deteriorates the achievable sum rate. However, when $N_s = L$, SPS behaves the same as PPS, because it ends up choosing all paths and cannot avoid correlation. When we set $N_s = L$, the likelihood of selecting paths with similar angular directions significantly increases, the rank of the beamforming matrices decreases and the achievable rate drops. This likelihood is more pronounced in the fair and rich scattering scenarios. In the fair scenario, SPS yields the best performance in the multiplexing range  $N_s/L = 0.125$ to $0.625$. Finally, the simulation results for the rich scattering scenario shown in Figure~\ref{fig:outer_rich} indicate a similar behavior to that observed in the fair scenario. The main difference is that PPS performs even worse. Overall, these results reveal that SPS yields the best performance when there is enough path diversity and the spatial multiplexing is not too aggressive. CME exhibits good robustness to strong spatial multiplexing. Although SPS performs better than CME in many scenarios, it is more computationally complex, especially when $M_t$ and $M_r$ are large.

Furthermore, figures~\ref{fig:outer_poor}--\ref{fig:outer_rich} provide valuable information on how to select the  transceiver parameters $M_r$ and $M_t$. Since $N_s = M_r = M_t$ in these experiments, we observe that $M_r/L = M_t/L$ can be set as large as $0.75$, $0.625$ and $0.375$ at poor, fair and rich scattering environments, respectively, for SPS. Larger ratios do not improve performance and may even deteriorate the achievable rate. Similar analysis can be done for CME and PPS. Note that we assumed $M_r = M_t$ for simplicity since the analysis becomes convoluted when $M_r \neq M_t$.

\subsection{Inner Layer Filters}

Recall that the inner filtering layer aims at tackling multi-user interference. Therefore, we conducted experiments to compare the interference robustness of the proposed inner layer schemes. We employed CME outer filtering motivated by the insights obtained from the outer layer simulation results.

\begin{figure}[t]
	\centering
%
%
\definecolor{mycolor1}{rgb}{0.00000,0.44700,0.74100}%
\definecolor{mycolor2}{rgb}{0.85000,0.32500,0.09800}%
\definecolor{mycolor3}{rgb}{0.92900,0.69400,0.12500}%
\definecolor{mycolor4}{rgb}{0.49400,0.18400,0.55600}%
\begin{tikzpicture}

\begin{axis}[%
width=2.5in,
height=2.5in,
scale only axis,
xmin=-20,
xmax=40,
xlabel style={font=\color{white!15!black}},
xlabel={SNR [dB]},
ymin=0,
ymax=80,
ylabel style={font=\color{white!15!black}},
ylabel={Achievable sum rate [bit/s/Hz]},
axis background/.style={fill=white},
xmajorgrids,
ymajorgrids,
legend style={at={(0.03,0.97)}, anchor=north west, legend cell align=left, align=left, draw=white!15!black}
]
\addplot [color=mycolor1, line width=1.0pt, mark= triangle,  mark size = 2pt]
  table[row sep=crcr]{%
-20	7.85282355572482\\
-15	12.1083119379266\\
-10	16.2526739462693\\
-5	19.79977106458\\
0	22.5167095631474\\
5	24.361054490396\\
10	25.4455958755565\\
15	25.9880290845634\\
20	26.2228588286543\\
25	26.3163074340032\\
30	26.3524744800371\\
35	26.3659245501046\\
40	26.3705830465381\\
};

\addplot [color=mycolor2, line width=1.0pt, mark=o, mark size = 2pt]
  table[row sep=crcr]{%
-20	1.27378510246058\\
-15	2.82663050563194\\
-10	5.36980172260152\\
-5	8.94561771146812\\
0	13.4644677230227\\
5	18.7455896695009\\
10	24.5707018502868\\
15	30.7458802801485\\
20	37.1305072874206\\
25	43.6336463354723\\
30	50.2005894547117\\
35	56.8001619120412\\
40	63.4180852337992\\
};

\addplot [color=green!60!black, line width=1.0pt, mark=square, mark size = 2pt]
  table[row sep=crcr]{%
-20	7.9564245144344\\
-15	12.5168042313545\\
-10	17.3906822429478\\
-5	22.2736096105771\\
0	27.0167341592655\\
5	31.5415013870853\\
10	35.8614451982972\\
15	40.0884126606172\\
20	44.380388170538\\
25	48.9091924133015\\
30	53.8155652713962\\
35	59.1470070187402\\
40	64.8636077813905\\
};

\addplot [color=mycolor3, line width=1.0pt, mark=x, mark size = 2pt]
  table[row sep=crcr]{%
-20	2.87893311366331\\
-15	5.65262053817159\\
-10	9.56856531596212\\
-5	14.4226652297536\\
0	19.9671066194038\\
5	25.9849183894281\\
10	32.2986794220489\\
15	38.7804533797202\\
20	45.352101173456\\
25	51.9677563136317\\
30	58.6017762261856\\
35	65.2424053849923\\
40	71.8852283787536\\
};

\node[rotate=45](metbd) at (axis cs:30, 45){\footnotesize MET-BD};
\node[rotate=45](bdmer) at (axis cs:31, 65){\footnotesize BD-MER};
\node(metmer) at (axis cs:30,22){\footnotesize MET-MER};
\node[rotate=32.5](metmmse) at (axis cs:0, 32){\footnotesize MET-MMSE};

\end{axis}
\end{tikzpicture}%
	\caption{Inner layer methods at poor scattering ($L=8$ paths), $M_t = M_r = 4$, $N_s = 1$ stream per user and $U=4$ UEs.}
	\label{fig:snr_4users}
\end{figure}
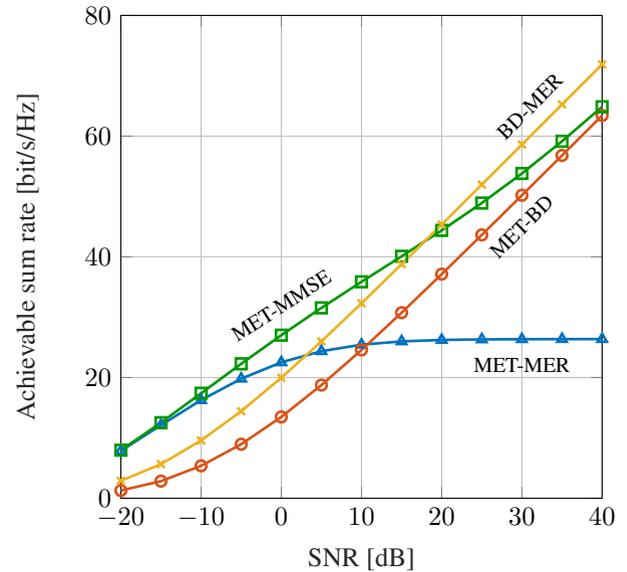

\begin{figure}[t]
	\centering
%
%
\definecolor{mycolor1}{rgb}{0.00000,0.44700,0.74100}%
\definecolor{mycolor2}{rgb}{0.85000,0.32500,0.09800}%
\definecolor{mycolor3}{rgb}{0.92900,0.69400,0.12500}%
\definecolor{mycolor4}{rgb}{0.49400,0.18400,0.55600}%
\begin{tikzpicture}

\begin{axis}[%
width=2.5in,
height=2.5in,
scale only axis,
xmin=-20,
xmax=40,
xlabel style={font=\color{white!15!black}},
xlabel={SNR [dB]},
ymin=0,
ymax=80,
ylabel style={font=\color{white!15!black}},
ylabel={Achievable sum rate [bit/s/Hz]},
axis background/.style={fill=white},
xmajorgrids,
ymajorgrids,
legend style={at={(0.03,0.97)}, anchor=north west, legend cell align=left, align=left, draw=white!15!black}
]
\addplot [color=mycolor1, line width=1.0pt, mark= triangle,  mark size = 2pt]
  table[row sep=crcr]{%
-20	13.2886515306397\\
-15	24.8271308795564\\
-10	37.1131513490988\\
-5	46.378581270998\\
0	51.6688167512542\\
5	54.1148727424628\\
10	55.0822395287292\\
15	55.4260092176472\\
20	55.5405426873799\\
25	55.5774886100461\\
30	55.589251562064\\
35	55.5929795538481\\
40	55.5941592789496\\
};


\addplot [color=green!60!black, line width=1.0pt, mark=square, mark size = 2pt]
  table[row sep=crcr]{%
-20	13.4074547109801\\
-15	25.5321056475626\\
-10	39.6746225092585\\
-5	52.5731811396617\\
0	62.6859708801521\\
5	69.8633931620207\\
10	74.4865534015833\\
15	77.1589673398296\\
20	78.5545321061236\\
25	79.236517526547\\
30	79.5640163996137\\
35	79.7231121561013\\
40	79.8012559121313\\
};


\node(metmer) at (axis cs:30,52){\footnotesize MET-MER};
\node(metmmse) at (axis cs:30, 75){\footnotesize MET-MMSE};

\end{axis}
\end{tikzpicture}%
	\caption{Inner layer methods at poor scattering ($L=8$ paths), $M_t = M_r = 4$, $N_s = 1$ stream per user and $U=32$ UEs.}
	\label{fig:snr_32users}
\end{figure}
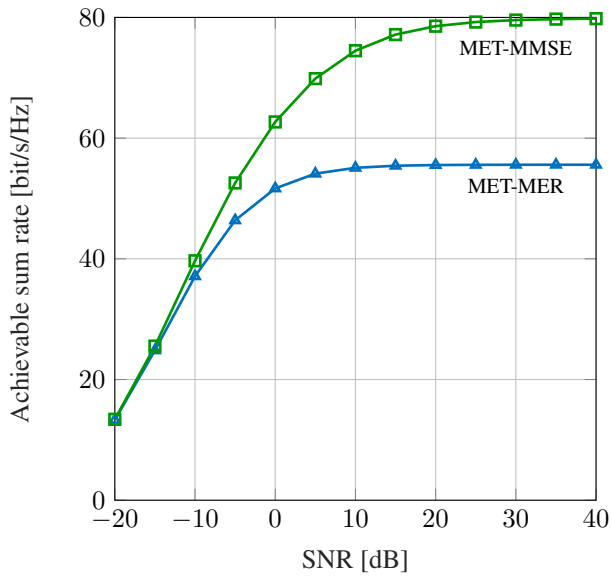

Let us begin the assessment of the inner layer filters by analyzing the achievable sum rate performance at the pessimistic (poor) propagation scenario. Figure~\ref{fig:snr_4users} shows the transceiver performance for a non-congested setup with $U=4$ UEs, $N_s = 1$ data stream per user and $M_t = M_r=4$. Since $UN_s = M_t = M_r$, BD/MMSE cancels the multi-user interference out, as expected. Also, all transceivers but MET-MER achieve the full degrees of freedom in the asymptotic SNR regime. What would happen in a congested scenario? In Figure~\ref{fig:snr_32users}, we consider $U=32$ UEs, $N_s=1$ data stream per user and $M_t = M_r = 4$. Note that this parameter setup gives $UN_s > M_t = M_r$, thus the BD conditions are not satisfied and the BD-based transceivers cannot be applied in this congested scenario. MET-MMSE works with this parameter setup, however, it is not able to completely reject the multi-user interference. As a result, the transceiver becomes interference-limited at high SNR. Nonetheless, we observe a reasonable performance at low SNR, e.g., MET-MMSE yields $63$ bit/s/Hz sum rate at $0$ dB SNR. This is because outer layer filtering already rejects some interference and the remainder is filtered by the inner layer. Figures~\ref{fig:snr_4users} and~\ref{fig:snr_32users} indicate that MET-MMSE and BD-MER yield the best performance in a non-congested scenario, while MET-MMSE and MET-MER are the preferred choices when the system becomes congested.

~

\begin{figure}[t]
	\centering
%
%
\definecolor{mycolor1}{rgb}{0.00000,0.44700,0.74100}%
\definecolor{mycolor2}{rgb}{0.85000,0.32500,0.09800}%
\definecolor{mycolor3}{rgb}{0.92900,0.69400,0.12500}%
\begin{tikzpicture}

\begin{axis}[%
width=2.5in,
height=2.5in,
scale only axis,
xmin=2,
xmax=64,
xlabel style={font=\color{white!15!black}},
xlabel={Number $U$ of UEs},
ymin=10,
ymax=90,
ylabel style={font=\color{white!15!black}},
ylabel={Achievable sum rate [bit/s/Hz]},
axis background/.style={fill=white},
xmajorgrids,
ymajorgrids,
xtick={4, 8, 16, 24, 32, 40, 48, 56, 64},
legend style={at={(0.03,0.97)}, anchor=north west, legend cell align=left, align=left, draw=white!15!black}
]
\addplot [color=mycolor1, line width=1.0pt, mark= triangle,  mark size = 2pt]
  table[row sep=crcr]{%
2	20.7953689520738\\
4	27.7770643401363\\
8	36.1145687503864\\
16	46.2785361039159\\
24	51.6554322539489\\
32	55.9439213131483\\
40	58.9069440455944\\
48	60.9414877978447\\
56	62.9162257926995\\
64	63.925347642412\\
};

\addplot [color=mycolor2, line width=1.0pt, mark=o, mark size = 2pt]
  table[row sep=crcr]{%
2	29.4161217696426\\
4	37.8610965627431\\
};

\addplot [color=green!60!black, line width=1.0pt, mark=square, mark size = 2pt]
  table[row sep=crcr]{%
2	29.5457135064732\\
4	45.3750744298984\\
8	58.7609335636175\\
16	71.9088990537502\\
24	76.3799659121898\\
32	78.8199496825159\\
40	80.3266984348587\\
48	81.0429334025825\\
56	81.47983068566\\
64	81.5806223471707\\
};

\addplot [color=mycolor3, line width=1.0pt, mark=x, mark size = 2pt]
table[row sep=crcr]{%
2	31.0322681878302\\
4	46.7716952291422\\
};

\node[rotate=45](metbd) at (axis cs:9, 42.5){\footnotesize MET-BD};
\node[rotate=45](bdmer) at (axis cs:8, 50){\footnotesize BD-MER};
\node(metmer) at (axis cs:52.5, 57){\footnotesize MET-MER};
\node(metmmse) at (axis cs:55, 77){\footnotesize MET-MMSE};

\end{axis}
\end{tikzpicture}%
	\caption{Inner layer methods at poor scattering ($L=8$ paths), $M_t = M_r = 4$, $N_s = 1$ stream per user and $\text{SNR}=20$ dB.}
	\label{fig:inner_poor}
\end{figure}
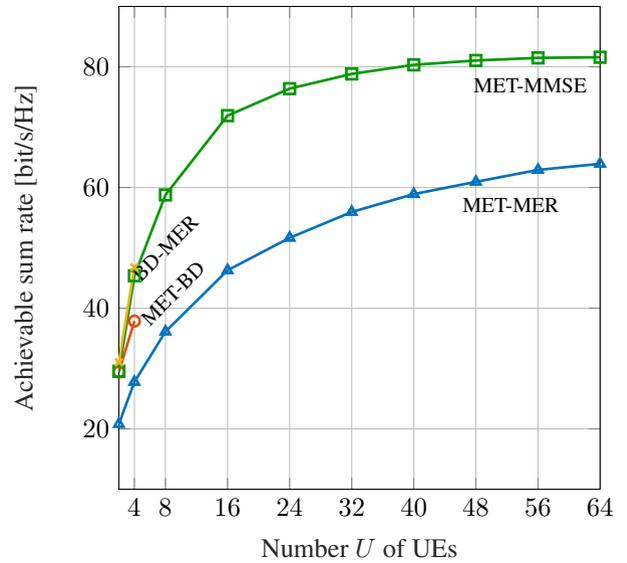

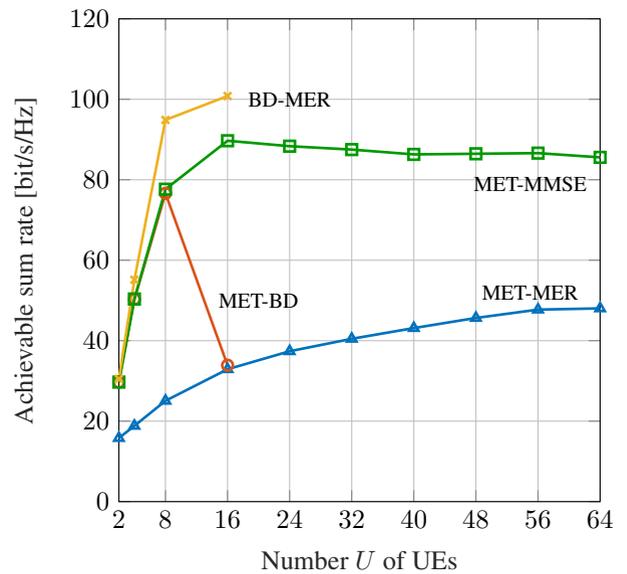
\begin{figure}[t]
	\centering
%
%
\definecolor{mycolor1}{rgb}{0.00000,0.44700,0.74100}%
\definecolor{mycolor2}{rgb}{0.85000,0.32500,0.09800}%
\definecolor{mycolor3}{rgb}{0.92900,0.69400,0.12500}%
\begin{tikzpicture}

\begin{axis}[%
width=2.5in,
height=2.5in,
scale only axis,
xmin=2,
xmax=64,
xlabel style={font=\color{white!15!black}},
xlabel={Number $U$ of UEs},
ymin=0,
ymax=120,
ylabel style={font=\color{white!15!black}},
ylabel={Achievable sum rate [bit/s/Hz]},
axis background/.style={fill=white},
xmajorgrids,
ymajorgrids,
xtick={2, 8, 16, 24, 32, 40, 48, 56, 64},
legend style={at={(0.03,0.97)}, anchor=north west, legend cell align=left, align=left, draw=white!15!black}
]
\addplot [color=mycolor1, line width=1.0pt, mark= triangle,  mark size = 2pt]
  table[row sep=crcr]{%
2	15.8022127731407\\
4	18.8329655255382\\
8	25.0567671709788\\
16	32.9031173331679\\
24	37.3772192626733\\
32	40.4664060631865\\
40	43.1364870377026\\
48	45.6487345793603\\
56	47.7232098181867\\
64	48.0266008727439\\
};

\addplot [color=mycolor2, line width=1.0pt, mark=o, mark size = 2pt]
  table[row sep=crcr]{%
2	29.7023460454444\\
4	50.3179178335579\\
8	76.6160612877132\\
16	33.8915438532\\
};

\addplot [color=green!60!black, line width=1.0pt, mark=square, mark size = 2pt]
  table[row sep=crcr]{%
2	29.7037460564905\\
4	50.3598522902768\\
8	77.6440030822474\\
16	89.6972552409993\\
24	88.3439071730382\\
32	87.501687935977\\
40	86.3027724550283\\
48	86.460207839737\\
56	86.6205330950833\\
64	85.5659296163815\\
};

\addplot [color=mycolor3, line width=1.0pt, mark=x, mark size = 2pt]
table[row sep=crcr]{%
2	30.4950641636868\\
4	55.1425353919531\\
8	94.8613427727491\\
16	100.812207616499\\
};

\node(metbd) at (axis cs:20,50){\footnotesize MET-BD};
\node(bdmer) at (axis cs:24,100){\footnotesize BD-MER};
\node(metmer) at (axis cs:55,52){\footnotesize MET-MER};
\node(metmmse) at (axis cs:55, 79){\footnotesize MET-MMSE};

\end{axis}
\end{tikzpicture}%
	\caption{Inner layer methods at fair scattering ($L=32$ paths) and $M_t = M_r = 16$, $N_s = 1$ stream per user and $\text{SNR}=20$ dB.}
	\label{fig:inner_fair}
\end{figure}

\begin{figure}[t]
	\centering
%
%
\definecolor{mycolor1}{rgb}{0.00000,0.44700,0.74100}%
\definecolor{mycolor2}{rgb}{0.85000,0.32500,0.09800}%
\definecolor{mycolor3}{rgb}{0.92900,0.69400,0.12500}%
\begin{tikzpicture}

\begin{axis}[%
width=2.5in,
height=2.5in,
scale only axis,
xmin=2,
xmax=64,
xlabel style={font=\color{white!15!black}},
xlabel={Number $U$ of UEs},
ymin=0,
ymax=210,
ylabel style={font=\color{white!15!black}},
ylabel={Achievable sum rate [bit/s/Hz]},
axis background/.style={fill=white},
xmajorgrids,
ymajorgrids,
xtick={2, 8, 16, 24, 32, 40, 48, 56, 64},
legend style={at={(0.03,0.97)}, anchor=north west, legend cell align=left, align=left, draw=white!15!black}
]
\addplot [color=mycolor1, line width=1.0pt, mark= triangle,  mark size = 2pt]
  table[row sep=crcr]{%
2	11.5461810223902\\
4	14.8884558739492\\
8	18.1716591488914\\
16	24.6067554357063\\
24	27.268059877945\\
32	30.2281444141746\\
40	31.875188369366\\
48	34.0940174237154\\
56	35.6044858314267\\
64	36.8514265927455\\
};

\addplot [color=mycolor2, line width=1.0pt, mark=o, mark size = 2pt]
  table[row sep=crcr]{%
2	28.76118642955\\
4	50.17325049369\\
8	80.070023223912\\
16	107.365848805794\\
24	77.5150043449868\\
32	5.83212800202529\\
};

\addplot [color=green!60!black, line width=1.0pt, mark=square, mark size = 2pt]
  table[row sep=crcr]{%
2	28.7617114481184\\
4	50.1817986250015\\
8	80.2093125190787\\
16	111.190011572985\\
24	112.346962500764\\
32	108.922753775399\\
40	103.202039579388\\
48	100.530195845473\\
56	98.2944364699944\\
64	96.3278795022424\\
};

\addplot [color=mycolor3, line width=1.0pt, mark=x, mark size = 2pt]
table[row sep=crcr]{%
2	29.9151358961801\\
4	54.7794392667755\\
8	96.8210948578562\\
16	164.992476260124\\
24	203.315285378419\\
32	138.090357429117\\
};

\node(metbd) at (axis cs:30, 75){\footnotesize MET-BD};
\node(metmer) at (axis cs:45, 40){\footnotesize MET-MER};
\node(bdmer) at (axis cs:34,170){\footnotesize BD-MER};
\node(metmmse) at (axis cs:55, 110){\footnotesize MET-MMSE};
\end{axis}
\end{tikzpicture}%
		\caption{Inner layer methods at rich scattering ($L=64$ paths) and $M_t = M_r = 32$, $N_s = 1$ stream per user and $\text{SNR}=20$ dB.}
	\label{fig:inner_rich}
\end{figure}
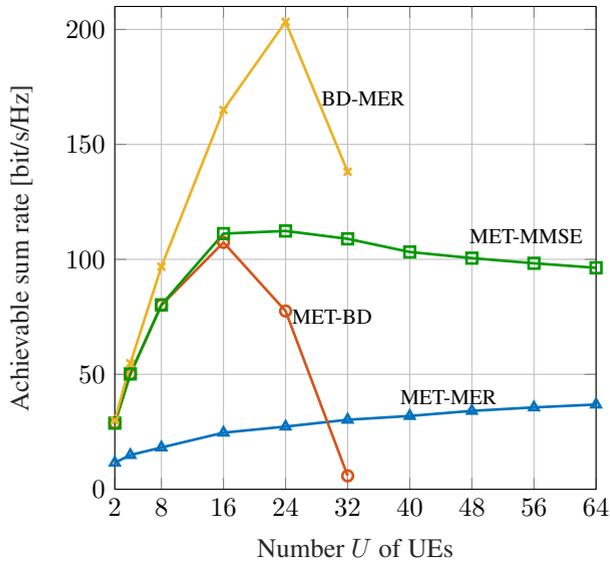

Figure~\ref{fig:snr_32users} motivates us to further study the robustness of the transceivers to UE congestion. To this end, we vary the number of UEs from $2$ to $64$ considering $N_s=1$ data stream per UE, $N_t = N_r = 64$ antennas, $M_r/L = M_t/L = 0.5$ and $\text{SNR} = 20\,\text{dB}$ for different scattering conditions in figures~\ref{fig:inner_poor},~\ref{fig:inner_fair} and~\ref{fig:inner_rich}. Figure~\ref{fig:inner_poor} shows the achievable sum rate performance for the poor scattering scenario. We observe that the BD-based transceivers do not perform well in this scenario, as they are capable to manage up to $4$ UEs. Note that BD-MER and MET-BD are plotted only when the BD condition $UN_s \leq M_t=M_r$ is satisfied. MET-MMSE and MET-MER, on the other hand, are not limited by this constraint and provide satisfactory results even when the system is overloaded. At $20$ dB SNR, the transceivers already have attained the rate saturation region, as we see in Figure~\ref{fig:snr_32users} when the system is congested. Therefore, these curves mainly compare how well the transceivers perform when the system becomes interference-limited. Figures \ref{fig:inner_fair} and \ref{fig:inner_rich} present the simulation results for the fair and rich scattering scenarios, respectively. As the environment offers more scatterers, the transceivers may operate with larger $M_t$ and $M_r$ and, consequently, more UEs can be served. Figures \ref{fig:inner_fair} and \ref{fig:inner_rich} reveal that BD-MER has  performance peaks at $16$ and $24$ UEs, respectively, which outperforms MET-MMSE for the given parameters. However, as $UN_s$ approaches $M_t$ and $M_r$, the performance of the BD-based transceivers deteriorates. In conclusion, spatial multiplexing in poor scattering scenarios should be carried out using either MET-MMSE or MET-MER since there are not enough degrees of freedom for BD to cancel the interference. When the propagation medium offers more scattering diversity, such as in the fair and rich scenarios, BD-MER becomes a reasonable choice as long $UN_s \leq M_t$. But even when this condition is not obeyed, MET-MMSE still provides proper results.

~
~

\subsection{Benchmarking}

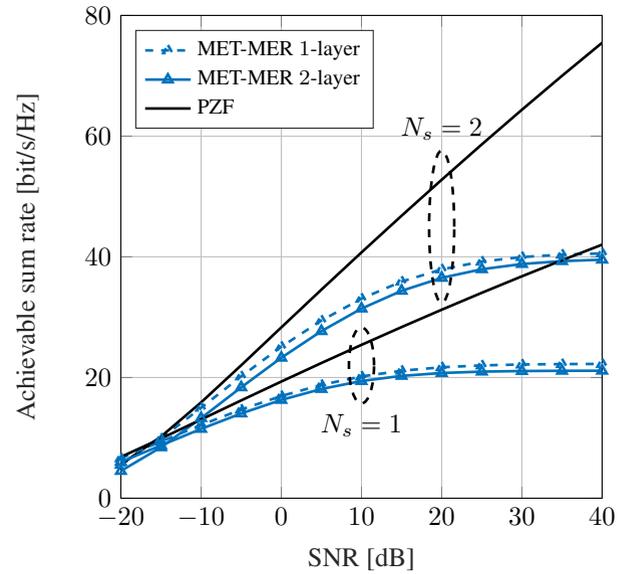
\begin{figure}[t]
	\centering
%
%
\begin{tikzpicture}

\definecolor{mycolor1}{rgb}{0.00000,0.44700,0.74100}%
\definecolor{mycolor2}{rgb}{0.85000,0.32500,0.09800}%
\definecolor{mycolor3}{rgb}{0.92900,0.69400,0.12500}%
\definecolor{mycolor4}{rgb}{0.49400,0.18400,0.55600}%

\begin{axis}[%
width=2.5in,
height=2.5in,
scale only axis,
xmin=-20,
xmax=40,
xlabel style={font=\color{white!15!black}},
xlabel={SNR [dB]},
ymin=0,
ymax=80,
ylabel style={font=\color{white!15!black}},
ylabel={Achievable sum rate [bit/s/Hz]},
axis background/.style={fill=white},
xmajorgrids,
ymajorgrids,
legend style={at={(0.03,0.97)}, font=\footnotesize, anchor=north west, legend cell align=left, align=left, draw=white!15!black}
]
\addplot [color=mycolor1, dashed, line width=1.0pt, mark= triangle,  mark size = 2pt]
  table[row sep=crcr]{%
-20	5.53564751739997\\
-15	9.88515922979433\\
-10	14.984125869977\\
-5	20.1904006695672\\
0	25.0925069070876\\
5	29.4393826389064\\
10	33.0685440787316\\
15	35.8965797539953\\
20	37.9166493393398\\
25	39.2139652725272\\
30	39.9634432838552\\
35	40.3628207729452\\
40	40.5665794648481\\
};
\addlegendentry{MET-MER $1$-layer}

\addplot [color=mycolor1, line width=1.0pt, mark= triangle,  mark size = 2pt]
  table[row sep=crcr]{%
-20	4.51653419660839\\
-15	8.43348705286549\\
-10	13.2707630595507\\
-5	18.3723000030508\\
0	23.2731224993716\\
5	27.6885244537031\\
10	31.4242895267594\\
15	34.3690779681059\\
20	36.5142929719482\\
25	37.944062863532\\
30	38.8086079657508\\
35	39.2846690754928\\
40	39.5258941163804\\
};
\addlegendentry{MET-MER $2$-layer}

\addplot [color=black, line width=1.0pt]
  table[row sep=crcr]{%
-20	5.58068856134152\\
-15	10.2053777239107\\
-10	15.8776894910731\\
-5	22.0253620520011\\
0	28.3056660716335\\
5	34.5615175458596\\
10	40.7328481638321\\
15	46.8043688762211\\
20	52.7726797413685\\
25	58.6311228220188\\
30	64.3737496093699\\
35	69.9939841575523\\
40	75.4742413320973\\
};
\addlegendentry{PZF}

\addplot [color=mycolor1, dashed, line width=1.0pt, mark= triangle,  mark size = 2pt]
  table[row sep=crcr]{%
-20	6.61534492256273\\
-15	9.3939611974193\\
-10	12.136248772944\\
-5	14.6884134646003\\
0	16.933966162244\\
5	18.7821092443863\\
10	20.1769908698596\\
15	21.124268492074\\
20	21.6986847525286\\
25	22.0090722256906\\
30	22.1603333848696\\
35	22.226766692751\\
40	22.2525414171742\\
};

\addplot [color=mycolor1, line width=1.0pt, mark= triangle,  mark size = 2pt]
  table[row sep=crcr]{%
-20	6.01702954839806\\
-15	8.7504605168305\\
-10	11.4864706427887\\
-5	14.0455839786031\\
0	16.2885251768323\\
5	18.099518796918\\
10	19.41867628715\\
15	20.2708812372312\\
20	20.747976746981\\
25	20.9773085688992\\
30	21.0754007982021\\
35	21.1137924084198\\
40	21.1275014169448\\
};

\addplot [color=black, line width=1.0pt]
  table[row sep=crcr]{%
-20	6.83763949838249\\
-15	9.88864212652407\\
-10	13.0440821452922\\
-5	16.204810255008\\
0	19.3305501046552\\
5	22.4029377684816\\
10	25.413712070457\\
15	28.3593176006106\\
20	31.2359034755503\\
25	34.0410503649041\\
30	36.7742943985384\\
35	39.4337666967952\\
40	42.014436399603\\
};

\node (ns1) at (axis cs:10 , 22) [ellipse, dashed, draw, label={below:$N_s=1$}, minimum width= 1mm, minimum height=10mm, line width = 1pt]  {};
\node (ns2) at (axis cs:20 , 45) [ellipse, dashed, draw, label={above:$N_s=2$}, minimum width= 1mm, minimum height=20mm, line width = 1pt]  {};

\end{axis}
\end{tikzpicture}%
	\caption{MET-MER benchmarking at poor scattering ($L=8$ paths), $M_t=M_r=4$, $U=2$ users and SNR $= 20$ dB.}
	\label{fig:bench_metmer}
\end{figure}

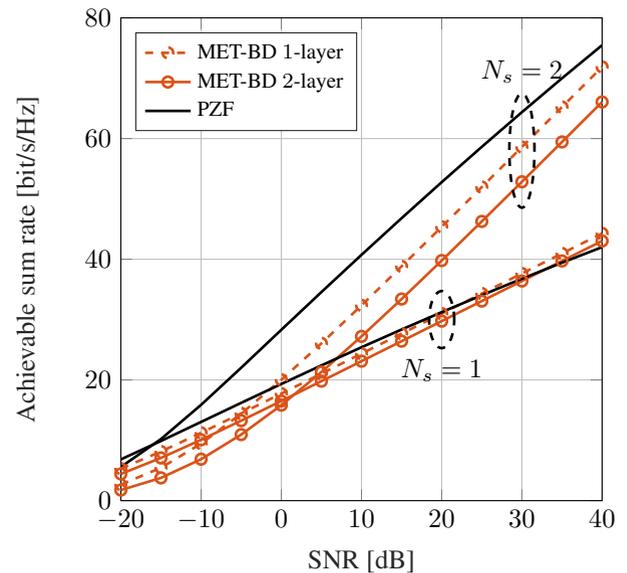
\begin{figure}[t]
	\centering
%
%
\begin{tikzpicture}

\definecolor{mycolor1}{rgb}{0.00000,0.44700,0.74100}%
\definecolor{mycolor2}{rgb}{0.85000,0.32500,0.09800}%
\definecolor{mycolor3}{rgb}{0.92900,0.69400,0.12500}%
\definecolor{mycolor4}{rgb}{0.49400,0.18400,0.55600}%

\begin{axis}[%
width=2.5in,
height=2.5in,
scale only axis,
xmin=-20,
xmax=40,
xlabel style={font=\color{white!15!black}},
xlabel={SNR [dB]},
ymin=0,
ymax=80,
ylabel style={font=\color{white!15!black}},
ylabel={Achievable sum rate [bit/s/Hz]},
axis background/.style={fill=white},
xmajorgrids,
ymajorgrids,
legend style={at={(0.03,0.97)}, font=\footnotesize, anchor=north west, legend cell align=left, align=left, draw=white!15!black}
]
\addplot [color=mycolor2, dashed, line width=1.0pt, mark=o, mark size = 2pt]
  table[row sep=crcr]{%
-20	2.65514949780706\\
-15	5.44477043776854\\
-10	9.45460083733854\\
-5	14.4128749575105\\
0	20.0218807312268\\
5	26.0562335135263\\
10	32.3586423758945\\
15	38.8207090768613\\
20	45.371373758237\\
25	51.9685231337708\\
30	58.5905242612012\\
35	65.2255381846062\\
40	71.866265549718\\
};
\addlegendentry{MET-BD $1$-layer}

\addplot [color=mycolor2, line width=1.0pt, mark=o, mark size = 2pt]
  table[row sep=crcr]{%
-20	1.75356743452965\\
-15	3.77092913446794\\
-10	6.86729398123895\\
-5	10.9453353067535\\
0	15.8248879971593\\
5	21.3108194377503\\
10	27.2194364449061\\
15	33.408894208845\\
20	39.7823928324768\\
25	46.273810467792\\
30	52.8371498909537\\
35	59.4415518998806\\
40	66.0677196233128\\
};
\addlegendentry{MET-BD $2$-layer}

\addplot [color=black, line width=1.0pt]
  table[row sep=crcr]{%
-20	5.58068856134152\\
-15	10.2053777239107\\
-10	15.8776894910731\\
-5	22.0253620520011\\
0	28.3056660716335\\
5	34.5615175458596\\
10	40.7328481638321\\
15	46.8043688762211\\
20	52.7726797413685\\
25	58.6311228220188\\
30	64.3737496093699\\
35	69.9939841575523\\
40	75.4742413320973\\
};
\addlegendentry{PZF}

\addplot [color=mycolor2, dashed, line width=1.0pt, mark=o, mark size = 2pt]
  table[row sep=crcr]{%
-20	5.30717509177192\\
-15	8.139554173667\\
-10	11.2348735460173\\
-5	14.454903301085\\
0	17.7330983218362\\
5	21.0379288771822\\
10	24.3538325395833\\
15	27.6737723892127\\
20	30.9950624393095\\
25	34.3167877998998\\
30	37.6386516866908\\
35	40.9605594673685\\
40	44.2824811373548\\
};

\addplot [color=mycolor2, line width=1.0pt, mark=o, mark size = 2pt]
  table[row sep=crcr]{%
-20	4.42560241323849\\
-15	7.08176950985793\\
-10	10.0827579425141\\
-5	13.2553497015442\\
0	16.5088282615642\\
5	19.8009394768081\\
10	23.1112936350121\\
15	26.429200579561\\
20	29.749812147285\\
25	33.0713190640484\\
30	36.3931134725516\\
35	39.7149992418648\\
40	43.0369139471945\\
};

\addplot [color=black, line width=1.0pt]
  table[row sep=crcr]{%
-20	6.83763949838249\\
-15	9.88864212652407\\
-10	13.0440821452922\\
-5	16.204810255008\\
0	19.3305501046552\\
5	22.4029377684816\\
10	25.413712070457\\
15	28.3593176006106\\
20	31.2359034755503\\
25	34.0410503649041\\
30	36.7742943985384\\
35	39.4337666967952\\
40	42.014436399603\\
};

\node (ns1) at (axis cs:20 , 30) [ellipse, dashed, draw, label={below:$N_s=1$}, minimum width= 1mm, minimum height=7.5mm, line width = 1pt]  {};
\node (ns2) at (axis cs:30 , 58) [ellipse, dashed, draw, label={above:$N_s=2$}, minimum width= 1mm, minimum height=15mm, line width = 1pt]  {};
\end{axis}
\end{tikzpicture}%
	\caption{MET-BD benchmarking at poor scattering ($L=8$ paths), $M_t=M_r=4$, $U=2$ users and SNR $= 20$ dB.}
	\label{fig:bench_metbd}
\end{figure}

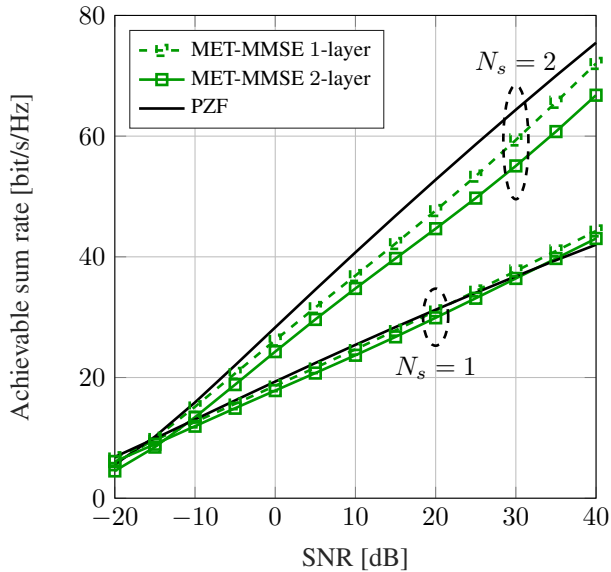
\begin{figure}[t]
	\centering
%
%
\begin{tikzpicture}

\definecolor{mycolor1}{rgb}{0.00000,0.44700,0.74100}%
\definecolor{mycolor2}{rgb}{0.85000,0.32500,0.09800}%
\definecolor{mycolor3}{rgb}{0.92900,0.69400,0.12500}%
\definecolor{mycolor4}{rgb}{0.49400,0.18400,0.55600}%

\begin{axis}[%
width=2.5in,
height=2.5in,
scale only axis,
xmin=-20,
xmax=40,
xlabel style={font=\color{white!15!black}},
xlabel={SNR [dB]},
ymin=0,
ymax=80,
ylabel style={font=\color{white!15!black}},
ylabel={Achievable sum rate [bit/s/Hz]},
axis background/.style={fill=white},
xmajorgrids,
ymajorgrids,
legend style={at={(0.03,0.97)}, font=\footnotesize, anchor=north west, legend cell align=left, align=left, draw=white!15!black}
]
\addplot [color=green!60!black, dashed, line width=1.0pt, mark=square, mark size = 2pt]
  table[row sep=crcr]{%
-20	5.54338515205984\\
-15	9.92880073892729\\
-10	15.1533929452244\\
-5	20.6759361819994\\
0	26.1963275416301\\
5	31.5943928683092\\
10	36.892093441575\\
15	42.2067535373235\\
20	47.6692475985293\\
25	53.3750106826543\\
30	59.3615100525716\\
35	65.6070500785927\\
40	72.0427558719287\\
};
\addlegendentry{MET-MMSE $1$-layer}

\addplot [color=green!60!black, line width=1.0pt, mark=square, mark size = 2pt]
  table[row sep=crcr]{%
-20	4.52274346043344\\
-15	8.47240746915923\\
-10	13.4285443463816\\
-5	18.8265605085894\\
0	24.2919552122062\\
5	29.6278354321477\\
10	34.7616435492428\\
15	39.731295828332\\
20	44.6646345603248\\
25	49.7286101332584\\
30	55.068131979503\\
35	60.757447998602\\
40	66.7799757896853\\
};
\addlegendentry{MET-MMSE $2$-layer}

\addplot [color=black, line width=1.0pt]
  table[row sep=crcr]{%
-20	5.58068856134152\\
-15	10.2053777239107\\
-10	15.8776894910731\\
-5	22.0253620520011\\
0	28.3056660716335\\
5	34.5615175458596\\
10	40.7328481638321\\
15	46.8043688762211\\
20	52.7726797413685\\
25	58.6311228220188\\
30	64.3737496093699\\
35	69.9939841575523\\
40	75.4742413320973\\
};
\addlegendentry{PZF}

\addplot [color=green!60!black, dashed, line width=1.0pt, mark=square, mark size = 2pt]
  table[row sep=crcr]{%
-20	6.70102627527187\\
-15	9.624900872179\\
-10	12.6372582758981\\
-5	15.6455712329819\\
0	18.6288933443794\\
5	21.6187783021691\\
10	24.6685344518562\\
15	27.814265155619\\
20	31.04831905537\\
25	34.3350002763038\\
30	37.6445794220182\\
35	40.9624522095232\\
40	44.2830815474552\\
};

\addplot [color=green!60!black, line width=1.0pt, mark=square, mark size = 2pt]
  table[row sep=crcr]{%
-20	6.08563064504487\\
-15	8.94957098188338\\
-10	11.9342489752321\\
-5	14.9105879030203\\
0	17.8363959988223\\
5	20.7400782032963\\
10	23.6885922728745\\
15	26.7383986079258\\
20	29.8972402189494\\
25	33.1363925698758\\
30	36.4201100076455\\
35	39.7252183698476\\
40	43.040439151723\\
};

\addplot [color=black, line width=1.0pt]
  table[row sep=crcr]{%
-20	6.83763949838249\\
-15	9.88864212652407\\
-10	13.0440821452922\\
-5	16.204810255008\\
0	19.3305501046552\\
5	22.4029377684816\\
10	25.413712070457\\
15	28.3593176006106\\
20	31.2359034755503\\
25	34.0410503649041\\
30	36.7742943985384\\
35	39.4337666967952\\
40	42.014436399603\\
};

\node (ns1) at (axis cs:20 , 30) [ellipse, dashed, draw, label={below:$N_s=1$}, minimum width= 1mm, minimum height=7.5mm, line width = 1pt]  {};
\node (ns2) at (axis cs:30 , 59) [ellipse, dashed, draw, label={above:$N_s=2$}, minimum width= 1mm, minimum height=15mm, line width = 1pt]  {};

\end{axis}
\end{tikzpicture}%
	\caption{MET-MMSE benchmarking at poor scattering ($L=8$ paths), $M_t=M_r=4$, $U=2$ users and SNR $= 20$ dB.}
	\label{fig:bench_metmmse}
\end{figure}

\begin{figure}[t]
	\centering
%
%
\begin{tikzpicture}

\definecolor{mycolor1}{rgb}{0.00000,0.44700,0.74100}%
\definecolor{mycolor2}{rgb}{0.85000,0.32500,0.09800}%
\definecolor{mycolor3}{rgb}{0.92900,0.69400,0.12500}%
\definecolor{mycolor4}{rgb}{0.49400,0.18400,0.55600}%

\begin{axis}[%
width=2.5in,
height=2.5in,
scale only axis,
xmin=-20,
xmax=40,
xlabel style={font=\color{white!15!black}},
xlabel={SNR [dB]},
ymin=0,
ymax=90,
ylabel style={font=\color{white!15!black}},
ylabel={Achievable sum rate [bit/s/Hz]},
axis background/.style={fill=white},
xmajorgrids,
ymajorgrids,
legend style={at={(0.03,0.97)}, font=\footnotesize, anchor=north west, legend cell align=left, align=left, draw=white!15!black}
]
\addplot [color=mycolor3, dashed, line width=1.0pt, mark=x, mark size = 2pt]
  table[row sep=crcr]{%
-20	5.56806468592547\\
-15	10.1889897893018\\
-10	15.8836241427601\\
-5	22.1145331843895\\
0	28.571771177942\\
5	35.1223668988326\\
10	41.7177631290023\\
15	48.3388627684563\\
20	54.9737082737556\\
25	61.6144290059312\\
30	68.2572592966408\\
35	74.9007874024012\\
40	81.5445394684796\\
};
\addlegendentry{BD-MER $1$-layer}

\addplot [color=mycolor3, line width=1.0pt, mark=x, mark size = 2pt]
  table[row sep=crcr]{%
-20	3.19176654671045\\
-15	6.00442863616017\\
-10	9.76849623749338\\
-5	14.2885940113141\\
0	19.3756699986652\\
5	24.8805462558807\\
10	30.6999439702768\\
15	36.7556368246484\\
20	42.9859300672592\\
25	49.3446440194447\\
30	55.7917429923257\\
35	62.2947449291042\\
40	68.8341887483511\\
};
\addlegendentry{BD-MER $2$-layer}

\addplot [color=black, line width=1.0pt]
  table[row sep=crcr]{%
-20	5.58068856134152\\
-15	10.2053777239107\\
-10	15.8776894910731\\
-5	22.0253620520011\\
0	28.3056660716335\\
5	34.5615175458596\\
10	40.7328481638321\\
15	46.8043688762211\\
20	52.7726797413685\\
25	58.6311228220188\\
30	64.3737496093699\\
35	69.9939841575523\\
40	75.4742413320973\\
};
\addlegendentry{PZF}

\addplot [color=mycolor3, dashed, line width=1.0pt, mark=x, mark size = 2pt]
  table[row sep=crcr]{%
-20	6.86303130935322\\
-15	9.97070734843417\\
-10	13.2203352117668\\
-5	16.5188846967226\\
0	19.8333659904301\\
5	23.1529337383766\\
10	26.4741148773351\\
15	29.7958067090686\\
20	33.1176600854891\\
25	36.4395645517728\\
30	39.7614851745846\\
35	43.0834109065887\\
40	46.4053382542662\\
};

\addplot [color=mycolor3, line width=1.0pt, mark=x, mark size = 2pt]
  table[row sep=crcr]{%
-20	5.38663710864168\\
-15	8.25206224756532\\
-10	11.3750075702046\\
-5	14.6100778756799\\
0	17.8942974498681\\
5	21.2013034005451\\
10	24.5179655862457\\
15	27.8381574180251\\
20	31.159528648535\\
25	34.4812798414687\\
30	37.8031519136283\\
35	41.1250622844055\\
40	44.4469847736194\\
};

\addplot [color=black, line width=1.0pt]
  table[row sep=crcr]{%
-20	6.83763949838249\\
-15	9.88864212652407\\
-10	13.0440821452922\\
-5	16.204810255008\\
0	19.3305501046552\\
5	22.4029377684816\\
10	25.413712070457\\
15	28.3593176006106\\
20	31.2359034755503\\
25	34.0410503649041\\
30	36.7742943985384\\
35	39.4337666967952\\
40	42.014436399603\\
};

\node (ns1) at (axis cs:20 , 32) [ellipse, dashed, draw, label={below:$N_s=1$}, minimum width= 1mm, minimum height=7.5mm, line width = 1pt]  {};
\node (ns2) at (axis cs:30 , 62) [ellipse, dashed, draw, label={above:$N_s=2$}, minimum width= 1mm, minimum height=15mm, line width = 1pt]  {};

\end{axis}
\end{tikzpicture}%
	\caption{BD-MER benchmarking at poor scattering ($L=8$ paths), $M_t=M_r=4$, $U=2$ users and SNR $= 20$ dB.}
	\label{fig:bench_bdmer}
\end{figure}
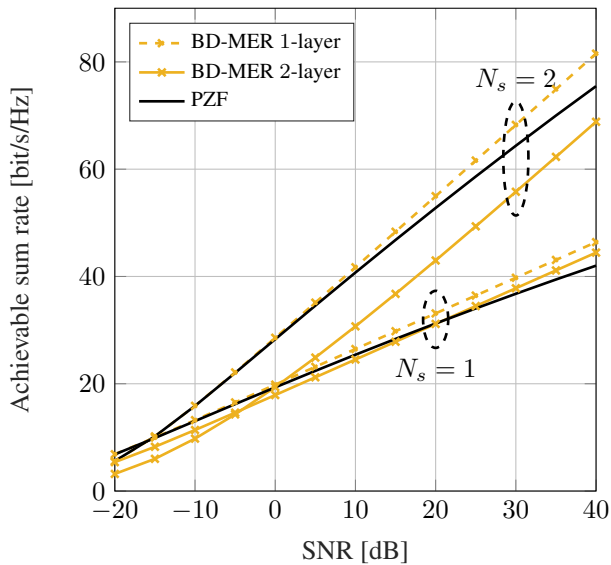

We benchmark the proposed transceiver to alternative schemes in this section. The first benchmark methods are the $1$-layer version of our proposed methods. They are based on the $(N_r \times N_t)$-dimensional channel matrix and they do not apply any outer-layer filter to form low-dimensional effective channels. The second benchmark method is the PZF solution proposed in~\cite{buzzi2018efficiency}. We assume perfect CSI for the $1$-layer and PZF benchmark methods. Figures~\ref{fig:bench_metmer}--\ref{fig:bench_bdmer} reproduce the benchmark results for the MET-MER, MET-BD, MET-MMSE and BD-MER transceivers, respectively. We consider the poor scattering scenario with $2$ UEs, $20$ dB SNR and $N_s = 1$ and $2$ data streams. Therefore, the BD condition is satisfied and the BD-based methods can be applied.

We observe that the $1$-layer strategy outperforms the proposed $2$-layer strategy in the achievable sum rate criterion in all benchmark results. This is expected because the $2$-layer solution is based on the concatenation of two filters, so any inaccuracy inserted by either outer or inner layer filter is sufficient to degrade the achievable performance relative to the $1$-layer version. However, the benchmark results indicate that these losses are negligible when only one data stream is transmitted. Among the proposed transceiver schemes, BD-MER exhibits the most important loss relative to its $1$-layer analogous at $N_s=2$ for the given parameters. PZF performs as well as our methods for $N_s=1$ data stream. However, we observe that PZF outperforms the proposed methods when the number of data streams is increased to $N_s=2$. PZF is a $1$-layer method, which does not rely on the concatenation of low-dimension filters, so its superior performance is expected in non-congested scenarios.

The benchmark methods exhibit, in general, larger data throughput than the proposed methods for the given simulation parameters. However, they are more computationally complex and CSI acquisition is unfeasible in practice due to the large dimensions of the associated CSI. Our methods, by contrast, have low computational complexity and practical CSI requirements, as discussed in sections~\ref{sec:csi} and~\ref{sec:comp}.

\section{Conclusion} \label{sec:conclusion}

We presented novel and practical transceiver schemes based on multi-layer filtering for double-sided massive MIMO systems. For the outer filtering layer, we compared a statistical approach (CME) to geometrical schemes (SPS and PPS). Simulation results show that SPS provides substantial gains over the naive PPS. Furthermore, it exhibits superior throughput to CME when spatial multiplexing is moderate, i.e., the number of data streams is roughly half the number of channel paths. However, the statistical approach offers good robustness to strong spatial multiplexing and can be less computationally complex than SPS. The choice between SPS and CME in practice amounts to the availability of either statistical or partial CSI. Regarding the inner filtering layer, MET-MMSE was found to be the most robust to different channel scattering conditions and multi-user interference, especially at low SNR. BD-MER provides the largest throughput for some specific scenarios with a fair amount of channel paths, which may not be practical in mmWave channels. For future work, we intend to investigate the proposed transceivers in some different application scenarios (multi-cell systems, vehicular communications, among others), to extend our methods to the broadband and multi-carrier scenarios~\cite{magueta2019hybrid} and to evaluate the effect of imperfect CSI on system performance.

\bibliographystyle{IEEEtran}
\bibliography{massivemimo.bib}

\begin{IEEEbiography}[{\includegraphics[width=1in,height=1.25in,clip,keepaspectratio]{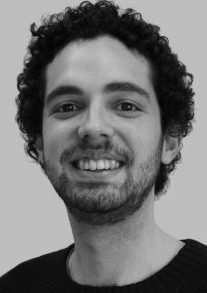}}]{Lucas N. Ribeiro}
	received his Bachelor degree in Teleinformatics Engineering from the Federal University of Cear\'a, Brazil, in 2014, and his Master degree in Informatics from the University of Nice Sophia Antipolis, France, in 2015. He received his Dr. Eng. degree in Teleinformatics Engineering from the Federal Univerisity of Cear\'a in 2019. He is currently a Postdoctoral researcher at Technische Universit\"at Ilmenau, Germany.
\end{IEEEbiography}
~
\begin{IEEEbiography}[{\includegraphics[width=1in,height=1.25in,clip,keepaspectratio]{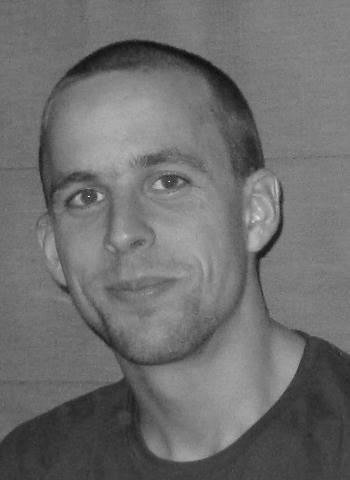}}]{Stefan Schwarz}
	received his Dr. techn. degree in telecommunications engineering in 2013 and his habilitation (post-doctoral degree) in the field of mobile communications in 2019, both from Technische Universität (TU) Wien. In 2010 he received the honorary price of the Austrian Minister
	of Science and Research and in 2014 he received the INiTS ICT award. From 2008 to 2015 he was working at the Institute of Telecommunications
	(ITC) of TU Wien as a University Assistant, conducting research and 4G and 5G mobile communication systems. Since 2016 he is heading the Christian Doppler Laboratory for Dependable Wireless Connectivity for the Society in Motion at ITC. He currently holds a tenure track position as Assistant Professor at TU Wien. His research interests are in wireless communications, channel measurements and characterization, link and system level simulations, and signal processing. 
\end{IEEEbiography}

\begin{IEEEbiography}[{\includegraphics[width=1in,height=1.25in,clip,keepaspectratio]{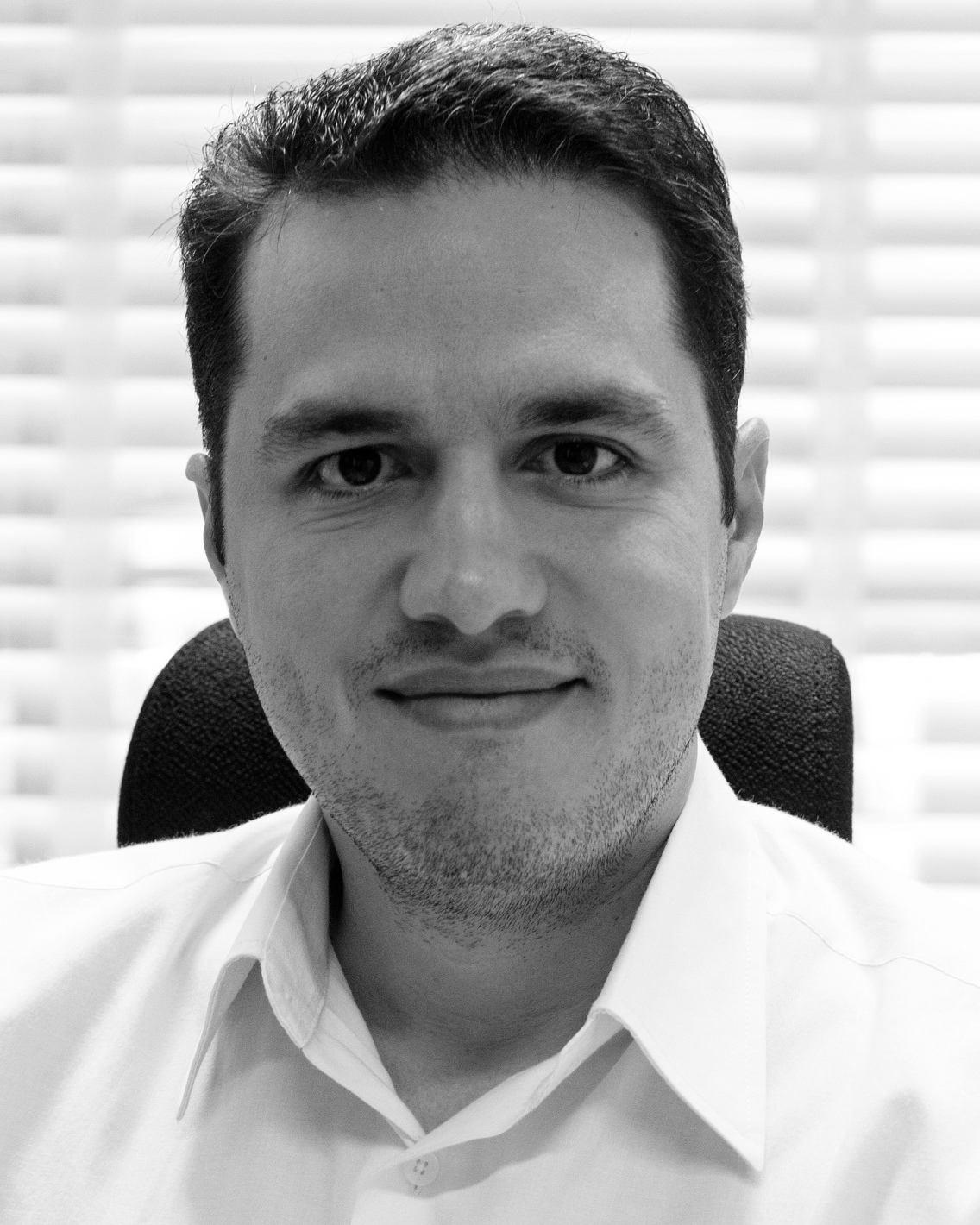}}]{Andr\'e L. F. de Almeida} is currently an Associate Professor with the Department of Teleinformatics Engineering of the Federal University of Ceará. He received a double Ph.D. degree in Sciences and Teleinformatics Engineering from the University of Nice, Sophia Antipolis, France, and the Federal University of Ceará, Fortaleza, Brazil, in 2007. During fall 2002, he was a visiting researcher at Ericsson Research Labs, Stockholm, Sweden. From 2007 to 2008, he held a one-year teaching position at the University of Nice Sophia Antipolis, France. In 2008, he was awarded a CAPES/COFECUB research fellowship with the I3S Laboratory, CNRS, France. He was awarded multiple times Visiting Professor positions at the University of Nice Sophia-Antipolis, France (2012, 2013, 2015, 2018, 2019). He has published over 200 refereed journal and conference papers. He served as an Associate Editor for the IEEE Transactions on Signal Processing (2012-2016). He currently serves as an Associate Editor for the IEEE Signal Processing Letters. He is a member of the Sensor Array and Multichannel (SAM) Technical Committee of the IEEE Signal Processing Society (SPS) and a member of the EURASIP Signal Processing for Multi-Sensor Systems Technical Area Committee (SPMuS – TAC). He was the General Co-Chair of the IEEE CAMSAP’2017 workshop and served as the Technical Co-Chair of the Symposium on “Tensor Methods for Signal Processing and Machine Learning” at IEEE GlobalSIP 2018 and IEEE GlobalSIP 2019. He also serves as the Technical Co-Chair of the IEEE SAM 2020 workshop, Hangzhou, China. He is a research fellow of the CNPq (the Brazilian National Council for Scientific and Technological Development). In January 2018, he was elected as an Affiliate Member of the Brazilian Academy of Sciences. He is a Senior Member of the IEEE. His research interests include the topics of channel estimation, sensor array processing, and multi-antenna systems. An important part of his research has been dedicated to multilinear algebra and tensor decompositions with applications to communications and signal processing.
\end{IEEEbiography}

\EOD

\end{document}